\documentclass[10pt,final, twocolumn]{IEEEtran}
\usepackage{amsmath,graphicx}
\usepackage{amssymb}
\usepackage{acronym}
\usepackage{pifont}

\usepackage[belowskip=-10pt,aboveskip=1.8pt]{caption}
\usepackage{standalone}
\usepackage{subcaption}
\usepackage{tikz}
\usepackage{url,cite}
\usepackage{verbatim}
\usepackage[bookmarks,colorlinks]{hyperref}
\usepackage{soul, xcolor}
\usepackage{enumitem}
\usepackage{mathtools}
\usepackage{hhline}
\usepackage{diagbox}
\usepackage{multirow}	
\usepackage{color, colortbl}	
\definecolor{Gray}{gray}{0.9}
\definecolor{LightCyan}{rgb}{0.88,1,1}

\usepackage[ruled,linesnumbered,vlined]{algorithm2e}
\SetKwInput{KwData}{\textbf{Init}} 

\newcommand{\myVec}[1]{{\boldsymbol{#1}}}
\newcommand{\myMat}[1]{{\boldsymbol{#1}}}
\newcommand{\mySet}[1]{\mathcal{#1}}

\acrodef{tbd}[TBD]{Track-Before-Detect}
\acrodef{hmm}[HMM]{hidden Markov model}
\acrodef{dl}[DL]{deep learning}
\acrodef{dnn}[DNN]{deep neural network}
\acrodef{snr}[SNR]{signal-to-noise ratio}
\acrodef{bs}[BS]{base station} 
\acrodef{cpu}[CPU]{centralized processing unit} 
\acrodef{mimo}[MIMO]{multiple-input multiple-output}
\acrodef{awgn}[AWGN]{additive white Gaussian noise} 
\acrodef{cpu}[CPU]{central processing unit} 
\acrodef{ml}[ML]{machine learning} 
\acrodef{mse}[MSE]{mean-squared error}
\acrodef{iot}[IOT]{Internet of Things}
\acrodef{rmse}[RMSE]{root mean squared error}
\acrodef{rmspe}[RMSPE]{root mean squared periodic error}
\acrodef{mmse}[MMSE]{{minimum mean-squared error}}
\acrodef{lmmse}[LMMSE]{{linear} MMSE}
\acrodef{mle}[MLE]{maximum likelihood estimation}
\acrodef{snr}[SNR]{signal-to-noise ratio}
\acrodef{sgd}[SGD]{stochastic gradient descent} 
\acrodef{dadmm}[D-ADMM]{distributed alternating direction method of multipliers}
\acrodef{sar}[SAR]{successive approximation register}
\acrodef{adc}[ADC]{analog-to-digital converter} 
\acrodef{dac}[DAC]{digital-to-analog converter} 
\acrodef{msb}[MSB]{most significant bit}
\acrodef{lsb}[LSB]{least significant bit}
\acrodef{pf}[PF]{particle filter}
\acrodef{apf}[APF]{auxiliary \ac{pf}}
\acrodef{pdf}[PDF]{probability density function}
\acrodef{tbd}[TBD]{track-before-detect}
\acrodef{map}[MAP]{maximum a posteriori}
\acrodef{ospa}[OSPA]{optimal subpattern assignment}
\acrodef{nn}[NN]{neural network}
\acrodef{fc}[FC]{fully connected}
\acrodef{sdp}[SDP]{scaled dot product} 
\acrodef{trapp}[TRAPP]{Target Resampling \ac{app}}
\acrodef{mtt}[MTT]{multi target tracking}
\acrodef{sa}[SA]{self attention}
\acrodef{lf}[LF]{learning flock}
\acrodef{lfpf}[LF-PF]{\ac{lf} augmented \ac{pf}}
\acrodef{lfapf}[LF-APF]{\ac{lf} augmented \ac{apf}}
\acrodef{apppf}[APP]{Auxiliary Parallel Partition \ac{pf}}
\acrodef{lfapppf}[LF-APP]{\ac{lf} augmented \ac{apppf}}
\acrodef{lfapppf}[LF-APP]{\ac{lf} augmented \ac{apppf}}
\acrodef{rnn}[RNN]{Recurrent Neural Network}
\acrodef{na}[NA]{Neural Augmented}
\acrodef{sis}[SIS]{Sequential Importance Sampling}
\acrodef{sispf}[SISPF]{\ac{sis} \ac{pf}}
\acrodef{lfsispf}[LF-SISPF]{\ac{lf} augmented \ac{sispf}}
\acrodef{pppf}[PPPF]{Progressive Proposal \ac{pf}}
\acrodef{urpf}[UrPF]{unrolled \ac{pf}}
\acrodef{lfurpf}[LF-UrPF]{\ac{lf} augmented \ac{urpf}}
\acrodef{fpm}[FPM]{floating point multiplications}

\usetikzlibrary{trees,arrows}

\setstcolor{blue}

\newcommand{\figsbasepath}{figs\_light}

\usepackage{contour}
\usepackage[normalem]{ulem}
\contourlength{0.8pt}

\title{
\vspace*{-2mm}
Learning Flock: Enhancing Sets of Particles for Multi~Sub-State Particle Filtering with Neural Augmentation}

\author{
\IEEEauthorblockN{Itai Nuri and Nir Shlezinger
\thanks{Parts of this work were accepted for presentation in the 2024 IEEE Signal Processing Advances in Wireless Communications (SPAWC) as the paper~\cite{nuri2024neural}.
The work was  supported by the Israel Innovation Authority. 
The authors are with the School of ECE, Ben-Gurion University of the Negev, Israel (email: itai5n@gmail.com; nirshl@bgu.ac.il).}
}}

\begin{document}

\maketitle
%
%

\begin{abstract} 

A leading family of algorithms for state estimation in  dynamic systems with multiple sub-states is based on \acp{pf}.  \acp{pf}  often struggle when operating under complex or approximated modelling (necessitating many particles) with low latency requirements (limiting the number of particles), as is typically the case in \ac{mtt}. 
In this work, we introduce a \ac{dnn} augmentation for \acp{pf} termed {\em \ac{lf}}. \ac{lf} learns to correct a particles-weights set, which we coin {\em flock}, based on the relationships between all sub-particles in the set itself, while disregarding the set acquisition procedure. 
Our proposed \ac{lf}, which can be readily incorporated into different \acp{pf} flow, is designed to facilitate rapid operation by  maintaining accuracy with a reduced number of particles. 
We introduce a dedicated training algorithm, allowing both supervised and unsupervised training, and yielding a module that 
supports a varying number of sub-states and  particles without necessitating re-training.
We experimentally show the improvements in performance, robustness, and latency of \ac{lf} augmentation for radar multi-target tracking, as well its ability to mitigate  the effect of a mismatched observation modelling.
We also compare and illustrate the advantages of \ac{lf} over a state-of-the-art \ac{dnn}-aided \ac{pf}, and demonstrate that \ac{lf} enhances both classic \acp{pf} as well as \ac{dnn}-based filters.
\end{abstract}

\acresetall

\section{Introduction}
\label{sec:intro}
 
The tracking of a time-evolving hidden state from noisy measurements is a fundamental signal processing task. \Acp{pf} are a family of algorithms that allow tracking in (possibly) non-linear and non-Gaussian dynamic systems~\cite{djuric2003particle}. \acp{pf} naturally support tracking a single state, as well as a state comprised  of multiple sub-states, i.e., \ac{mtt}.   Such filters are widely popular, and are utilized   in many areas, ranging from robotics~\cite{thrun2002particle}, {through communication~\cite{djuric2002applications}, and to}  positioning~\cite{gustafsson2002particle} and radar  tracking ~\cite{boers2004multitarget,ubeda2017adaptive, ito2020multi}.  

\acp{pf} employ Monte Carlo simulation to iteratively update a set of samples coined {\em particles}, that represent sampling points of a continuous \ac{pdf}, and their corresponding weights, which represent their relative correctness~\cite[Ch. 12]{durbin2012time}.
Their update is done according to a sequence of observations and the statistical modelling of the state evolution and the measurements, via different sampling-based stochastic procedures. 
While this operation enables the tracking of complex distributions, it also gives rise to $(i)$ increased complexity due to the need to maintain a large number of particles; and $(ii)$ sensitivity to lacking or mismatched knowledge of the related models, that affects the update of the particles and may limit application in real-time systems~\cite{buzzi2005track}. 
To tackle these challenges, \acp{pf} often employ, e.g., adaptation of the number of particles~\cite{grisetti2005improving, closas2011particle}, robust designs~\cite{xu2013robust,fisch2021innovative},  or intricate sampling and resampling realizations \cite{li2015resampling,doucet2000sequential, piavanini2023annealed, ubeda2017adaptive, ito2020multi}, typically at the cost of increased complexity and latency and/or reduced performance. 

Over the last decade, \acp{dnn} have emerged as powerful data-driven tools, that allow learning complex and abstract mappings from data~\cite{Goodfellow_Deep_Learning}. The dramatic success of deep learning in domains such as computer vision and natural language processing has also lead to a growing interest in its combination with classic signal processing algorithms via model-based deep learning~\cite{shlezinger2020model, shlezinger2022model,shlezinger2023model}. For tracking in dynamic systems, \ac{dnn}-aided implementations of Kalman-type filters, that are typically suitable for tracking a single state in a Gaussian dynamic system, were studied in~\cite{revach2022kalmannet,revach2022unsupervised,buchnik2023latent,ghosh2023danse,ni2022rtsnet,choi2023split}.

In the context of \acp{pf}, various approaches were proposed in the literature to incorporate deep learning \cite{chen2021differentiable, jonschkowski2018differentiable,ma2020particle,gama2022unrolling, gama2023unsupervised, cox2024end, piavanini2024deep,xia2021multi}. 
In the last few years, arguably the most common approach uses \acp{dnn} to learn the sampling distribution, from which the particles are then individually sampled~\cite{chen2021differentiable,  gama2022unrolling, ma2020particle, jonschkowski2018differentiable,gama2023unsupervised, cox2024end, piavanini2024deep}. This can be done via supervised \cite{chen2021differentiable, jonschkowski2018differentiable,ma2020particle} and unsupervised \cite{gama2022unrolling,gama2023unsupervised,cox2024end,piavanini2024deep} learning. The latter is typically done by imposing a specific distribution model on the signals~\cite{gama2022unrolling,gama2023unsupervised,cox2024end}, thus  inducing limitations when this assumption is violated. Alternatively, one can learn by mimicking each particle of another \ac{pf} with the same number of particles that is considered accurate~\cite{piavanini2024deep}, and thus be bounded by its reference performance, which is often restrictive when operating with a limited number of particles. 
 In multi sub-state settings,  \acp{dnn} were used mainly as preprocessing~\cite{xia2021multi}, less so as an integral part of the \ac{pf} flow.
While the above existing neural augmentations were shown to enhance \acp{pf}, their design is typically tailored  to a specific task, limiting transferability to other filters. Moreover, they operate in a per-particle manner, thus do not fully leverage the relationship between particles,  e.g.  in order to disperse clusters of particles or to align outliers.  This motivates the design of a generic learning-aided improvement to \acp{pf} that induces a flexible collective distribution between all particles, by compactly utilizing available models knowledge  in the form of the pertinent \ac{pf}, and by leveraging data to exploit shared information between particles.

In this work we present a novel approach of augmenting multi sub-state \acp{pf} with \ac{dnn}, coined {\em \ac{lf}}. 
Our algorithm is based on the insight that the core challenges of \acp{pf} with a limited number of particles can be tackled by learning from data to {\em jointly} correct particles and their weights, that are otherwise {\em independent},  at specific points in the \ac{pf} flow, so that the  particles and the weights at the end of each \ac{pf} iteration better reflect the state probability distribution. 
Accordingly, our proposed neural augmentation acts as a correction term to the particles and their weights, by adjusting the set collectively. The resulting augmentation is designed to be readily {\em transferable}, such that the \ac{lf} module can be in fact integrated into different \ac{pf} algorithms, and even combined in a {\em complementary} fashion with alternative \ac{dnn}-aided \acp{pf}.

In particular, we design a \ac{dnn} architecture for jointly processing varying number of particles with varying number of sub-states. We identify a core permutation equivariance of particle-weight pairs, recognizing that the induced distribution is invariant to their ordering. Accordingly, we design our \ac{lf} \ac{dnn} architecture to employ dedicated embedding modules to handle the invariance of sub-state indexing, and incorporate compact, trainable self-attention modules~\cite{vaswani2017attention} to account for the fact that the filter  is invariant to the particles' order.  

We introduce an  algorithm that boosts transferability by training the \ac{lf} module  as a form of generative learning~\cite{shlezinger2022discriminative}. This is achieved using a dedicated loss function that evaluates a set of particles and weights, in $(i)$ state recovery; and $(ii)$ the similarity of the particles spread to a desired target pattern. 
We propose both  supervised and unsupervised learning schemes. The latter is realized while boosting operation with a limited number of particles by utilizing an accurate teacher \ac{pf}, possibly with a large number of particles, as a form of knowledge distillation~\cite{gou2021knowledge}. Instead of distilling by aiming to mimic the particles and their weights, our loss evaluates the similarity in the particles' patterns induced by the the two \acp{pf},  boosting exploitation of cross-particle relationships without requiring a large number of particles.
 %
 
We numerically exemplify the gains of our proposed \ac{lf} for augmenting \acp{pf} in different tasks. We show that it can enhance different \acp{pf} for different synthetic data distributions. Moreover, we show that it can not only outperform alternative \ac{dnn}-aided \acp{pf}, but can also be integrated and enhance the operation of such algorithms. We also evaluate the \ac{lf} for radar \ac{mtt}, augmenting   \aclp{apf}~\cite{pitt1999filtering} in single and multiple radar target tracking settings~\cite{ubeda2017adaptive}. We show systematic improvements when augmenting our \ac{lf} module in performance, latency, and in overcoming modelling mismatches.

The rest of this paper is organized as follows: Section~\ref{sec:System Model} reviews the system model and recalls \ac{pf} basics; Section~\ref{sec:LF} derives the \ac{lfpf}, introducing its rationale, architecture and training algorithm. Section~\ref{sec:Experimental Study} presents an experimental study. Finally, Section~\ref{sec:conclusions} concludes the paper.

 Throughout the paper, we use boldface letters for vectors, e.g., ${\myVec{x}}$. 
 Upper-cased boldface letters denote matrices, e.g., $\myMat{X}$. 
 Calligraphic letters, such as $\mySet{X}$, are used for sets, with $|\mySet{X}|$ being the cardinality of $\mySet{X}$. 
 
\section{System Model and Preliminaries}
\label{sec:System Model}
This section reviews the system model along with necessary preliminaries. We first present the  signal model in Subsection~\ref{ssec:overallsystem}, based on which we formulate the tracking problem in Subsection~\ref{ssec:Problem}. Then, on Subsection~\ref{ssec:Paticle} we review \acp{pf} in general, and lay out the key principles that we base upon our \ac{lf} algorithm, detailed in Section~\ref{sec:LF}.

\subsection{Signal Model}
\label{ssec:overallsystem}  

We consider a dynamic system formulated as a continuous-valued state-space model in discrete-time. The state vector at time $k$, denoted  $\myVec{x}^k\in\mathbb{R}^{d_p}$, describes the states of $t$ dynamic sub-states. Accordingly, $\myVec{x}^k$  is comprised of  $t$ sub-vectors $\{\myVec{x}_{j}^{k}\}_{j=1}^{t}$ of size $d_{sp}\times 1$, with  $d_p=t\cdot d_{sp}$,  such that $\myVec{x}^{k}=\left[\myVec{x}_{1}^{k},\myVec{x}_{2}^{k},..,\myVec{x}_{j}^{k},..,\myVec{x}_{t}^{k}\right]$. For instance, different sub-states can represent different targets in \ac{mtt}~\cite{vo2015multitarget}.

Each sub-state vector  evolves in time 
 independently of the other sub-states, obeying a first-order Markov process. Specifically, we write the $j$th sub-state vector as $\myVec{x}^{0:k}_j = \{\myVec{x}_j^0,\ldots, \myVec{x}^k_j\}$, and assume that it obeys a motion model such that its conditional \ac{pdf} satisfies
\begin{equation}
\label{eqn:StateEvolutionTarg}
    p\left(\myVec{x}^k_j | \myVec{x}^{0:k-1}_j\right) = p\left(\myVec{x}^k_j |\myVec{x}^{k-1}_j\right),\qquad  \forall j\in \{1,\ldots,t\}.
\end{equation}
Accordingly, by writing  $\myVec{x}^{0:k}= \{\myVec{x}^0,\ldots, \myVec{x}^k\}$, the overall state  evolution \ac{pdf} satisfies 
\begin{equation}
\label{eqn:StateEvolution}
    p\left(\myVec{x}^k | \myVec{x}^{0:k-1}\right) = p\left(\myVec{x}^k |\myVec{x}^{k-1}\right),
\end{equation}  where $p\left(\myVec{x}^k |\myVec{x}^{k-1}\right)=\prod_{j=1}^{t}{p\left(\myVec{x}^k_j |\myVec{x}^{k-1}_j\right)}$. 

 At each time-step $k$, the state is observed via  noisy, possibly non-linear measurements,  denoted  $\myVec{z}^k\in\mathbb{R}^{d_m}$.  The relationship between the observations, i.e., the sensory data, and the state are reflected in the sensor model  \ac{pdf}
\begin{equation}
    \label{eqn:Measurements}
    p\left(\myVec{z}^k | \myVec{x}^{0:k}\right) = p\left(\myVec{z}^k |\myVec{x}^{k}\right).
\end{equation}
We particularly focus in settings where the observation model is invariant to the indexing of the sub-states. Mathematically, this indicates that the observation model in \eqref{eqn:Measurements} is invariant to replacing $\myVec{x}^{k}= \left[\myVec{x}^k_1,\ldots, \myVec{x}^k_t\right]$  with $\left[\myVec{x}^k_{j_1},\ldots, \myVec{x}^k_{j_t}\right]$ for any permutation $j_1,\ldots, j_t$ of $1,\ldots, t$.





  

\subsection{Problem Formulation}
\label{ssec:Problem}
Our objective is to design a system for recovering the state vector from all available data. Accordingly, for each time instance $k$, we are interested in providing an estimate of $\myVec{x}^k$ based on all past and current measurements, i.e., $\myVec{z}^{1:k}$. To focus on tracking tasks, we assume that the initial state is known.  

%
We  particularly consider multi sub-state tracking subject to 
the following requirements:
\begin{enumerate}[label={R\arabic*}]
    \item \label{itm:dist} The state evolution \ac{pdf} \eqref{eqn:StateEvolution} and measurement model  \eqref{eqn:Measurements} can be non-Gaussian.
    \item \label{itm:mismatch} The measurement model  \eqref{eqn:Measurements} is given yet may be mismatched.
    \item \label{itm:latency} The system must operate in real-time with low latency. 
    \item \label{itm:NumTargets}  The number of sub-states $t$ can take different values. Yet, for any given set of observations, we assume that it is known.
\end{enumerate}

To cope with the above challenges, we consider settings where we have access to data for design purposes. We consider two possible settings: 
\begin{enumerate}[label={S\arabic*}]
    \item \label{itm:settings1} 
    {\em Unsupervised settings} - here the data is comprised of a set of $n_t$ series of measurements along $\kappa$ steps long trajectories, denoted
    $ \mySet{D} = \left\{\left(\myVec{z}^{1:\kappa}\right) \right\},|\mySet{D}|=n_t$.
    \item \label{itm:settings2} 
    {\em Supervised settings} - here we also have access to the true states corresponding to the measurements as well as the measurements, i.e., 
   $ \mySet{D} = \left\{\left(\myVec{z}^{1:\kappa},\bar{\myVec{x}}^{1:\kappa}\right) \right\},|\mySet{D}|=n_t$.

\end{enumerate}

\subsection{Particle Filters}
\label{ssec:Paticle} 
\Acp{pf} are a popular family of algorithms suitable for tracking in non-Gaussian dynamics~\cite{djuric2003particle,bergman2001optimal} (thus meeting \ref{itm:dist}).  These Monte-Carlo algorithms~\cite{durbin2012time} track  a representation of  the posterior  $p(\myVec{x}^k|\myVec{z}^{1:k})$, by approximating its Bayesian recursive formulation which holds under the Markovian model in \eqref{eqn:StateEvolutionTarg} and \eqref{eqn:StateEvolution}.  Specifically, by Bayes' law it holds that~\cite{arulampalam2002tutorial}
\begin{equation}
     \label{eqn:IterativeBayes}         p(\myVec{x}^k|\myVec{z}^{1:k})=\frac{p(\myVec{z}^k|\myVec{x}^{k})p(\myVec{x}^k|\myVec{z}^{1:k-1})}{p(\myVec{z}^k|\myVec{z}^{1:k-1})}.
\end{equation}
Under model assumptions \eqref{eqn:StateEvolutionTarg}-\eqref{eqn:StateEvolution}, the posterior at time $k$ in \eqref{eqn:IterativeBayes} can be related to the posterior at time $k-1$ based on Chapman–Kolmogorov equation, i.e.,
 \begin{equation}
          \label{eqn:ChCoEq2}
         p(\myVec{x}^k|\myVec{z}^{1:k-1})=\int{p(\myVec{x}^k|\myVec{x}^{k-1})p(\myVec{x}^{k-1}|\myVec{z}^{1:k-1})d\myVec{x}^{k-1}}
   \end{equation}
and on
\begin{equation}
\label{eqn:ConditionalIntegral2}
         p(\myVec{z}^k|\myVec{z}^{1:k-1})=\int{p(\myVec{z}^k|\myVec{x}^{k})p(\myVec{x}^{k}|\myVec{z}^{1:k-1})d\myVec{x}^{k}},
    \end{equation}     
 turning \eqref{eqn:IterativeBayes} into a recursive update rule.

    Direct recursive updating of the posterior based on \eqref{eqn:IterativeBayes}-\eqref{eqn:ConditionalIntegral2} is often challenging or infeasible to compute. \acp{pf} approximate it 
using a set of $N$ particles ${\{\myVec{x}^k_i\}}_{i=1}^N$ and their weights ${\{w^k_i\}}_{i=1}^N$. Each particle represents a hypothesis on the system's state, and the weights indicate their  trajectory's relative accuracy. 
As the state vector represents $t$ sub-states, each particle  $\myVec{x}_i^k$ consists of $t$ sub-particles $\{\myVec{x}^k_{j,i}\}_{j=1}^t$ that share the same weight $w^k_i$.
The posterior  \ac{pdf} can be estimated as a weighted sum of  kernel functions $\{K_{j,i}(\cdot)\}_{j,i=1}^{t,N}$ evaluated at the sub-particles~\cite{arulampalam2002tutorial} via 
    \begin{align}
    \label{eqn:posterior}
         \hat{p} \left(\myVec{x}^k|\myVec{z}^{1:k}\right)=\prod_{j=1}^{t}{\sum_{i=1}^{N}{w^k_i K_{j,i}\left(\myVec{x}^{k}_j-{\myVec{x}}_{j,i}^{k}\right)}}.
    \end{align} 
By using an adequate number of particles, \acp{pf} can approximate any probability distribution. 
The state is then  recovered as a weighted average of the particles, i.e.,  
    \begin{align}
    \label{eqn:Estimate}
         \hat{\myVec{x}}^k&=\sum_{i=1}^{N}{w^k_i\myVec{x}^k_{i}}=
         \left[\sum_{i=1}^{N}{w^k_i\myVec{x}^k_{1,i}},\ldots,\sum_{i=1}^{N}{w^k_i\myVec{x}^k_{t,i}}\right].
    \end{align} 
Based on the principle of the {\em importance sampling} ~\cite{bergman1999recursive},
Algorithm~\ref{alg:PF} describes a generic  \ac{pf} iteration, which is comprised of two main stages: $(i)$ sampling current time-step particles $\{\myVec{x}_i^k\}_{i=1}^N$ using the {\em importance density} ~\cite{bergman1999recursive} $q\left(\myVec{x}^k\right)$ (Steps~\ref{stp:for1}-\ref{stp:advance}), and $(ii)$ update the weights $\{w_i^{k}\}_{i=1}^N$ according to the sampled density $q(\myVec{x}^k)$ and real distribution $\pi(\myVec{x}^k)$ to represent the relative accuracy of each trajectory with the new particles   (Steps~\ref{stp:for2}-\ref{stp:reweight}). 
By the Markovian nature of the signal, the real distribution satisfies $\pi(\myVec{x}^{0:k}|\myVec{z}^{1:k})\propto \pi(\myVec{z}^{k}|\myVec{x}^{k})\pi(\myVec{x}^{k}|\myVec{x}^{k-1})\pi(\myVec{x}^{0:k-1}|\myVec{z}^{1:k-1})$. 
The sampling distribution $ q\left(\cdot\right)$ is typically chosen to be of a decomposable form~\cite{djuric2003particle}, such that $q(\myVec{x}^{0:k}|\myVec{z}^{1:k})=q(\myVec{x}^k|\myVec{x}^{0:k-1},\myVec{z}^{1:k})q(\myVec{x}^{0:k-1}|\myVec{z}^{1:k-1})$.  This choice of $q(\cdot)$ and $\pi(\cdot)$  leads to a generic  weight update rule (Step \ref{stp:reweight}) in which
 \begin{equation}
     w^k_i \propto     \frac{\pi(\myVec{z}^{k}|\myVec{x}_i^{k})\pi(\myVec{x}_i^{k}|\myVec{x}_i^{k-1})}{q(\myVec{x}_i^{k}|\myVec{x}_i^{0:k-1},\myVec{z}^{1:k})}w^{k-1}_i.
 \end{equation}

\begin{algorithm}
    \caption{{A generic} PF iteration on time $k$}
    \label{alg:PF} 
    \SetKwInOut{Input}{Input}  
    \Input{Particles-weights set $\{\myVec{x}_i^{k-1},w_i^{k-1}\}_{i=1}^{N}$, measurements $\myVec{z}^{k}$}  
    {
        \For{$i = 1, \ldots, N$}{
        \label{stp:for1}
             Sample updated particles: \
             $\myVec{x}^k_i\sim q\left(\myVec{x}^k|\cdot\right)$\;
              \label{stp:advance}
        }
        \For{$i = 1, \ldots, N$}{
        \label{stp:for2}
            Update weights: \        $w_i^k\leftarrow{{\pi\left(\myVec{x}^k_i|\cdot\right)}/{q\left(\myVec{x}^k_i|\cdot\right)}}$\label{stp:reweight}\;
        }
        \For{$i = 1, \ldots, N$}{
        \label{stp:for3}
            Normalize: $w^k_i\leftarrow {w^k_i}/{\sum_{j=1}^{N}w^k_j}$ \; 
            \label{stp:normalize}
        }
        \SetKwInOut{Output}{Output} 
        \label{stp:output}
    \Output{Particles-weights set $\{\myVec{x}_i^{k},w_i^{k}\}_{i=1}^{N}$}
    }
\end{algorithm}

For increased accuracy, \acp{pf} require more particles, and this requirement grows dramatically with the state dimension \cite{rebeschini2015can}. The Number of particles greatly affects  the complexity of \acp{pf},
performance however is not necessarily dictated by the number of particles $N$, and is typically influenced by the {\em effective number of particles}, defined as~\cite{bergman1999recursive} 

     \begin{equation}
         {N_{\rm eff}}= \frac{N_{}}{1+ {\rm Var}\left(\{w^k_i\}_{i=1}^N\right)}.
    \end{equation}

Specifically, small ${N_{\rm eff}}$ indicates a phenomenon referred to as   {\em particles degeneracy}~\cite{djuric2003particle,arulampalam2002tutorial},  where the posterior is poorly estimated due to being dominated by a small portion of the overall particles. 
Having a large $N_{\rm eff}$  is typically a desired property of \acp{pf}. Nonetheless, on its own it does not  ensure a good representation of the posterior, as one may still encounter  {\em sample impoverishment}~\cite{djuric2003particle,arulampalam2002tutorial} (or particle collapse), where many particles converge to the same location. 
Applying particles resampling ~\cite{doucet2000sequential,djuric2003particle,li2015resampling} when $N_{\rm eff}$  goes below a threshold $N_{\rm th}$  tackles both phenomena to a degree.

        


\acp{pf} rely on stochastic procedures,  and are prone to producing outliers or deformations, particularly when the number of particles $N$ is small. However, the handling of a large number of particles, even if adaptive~\cite{straka2009survey}, comes with additional overhead and limit real time applications imperiling~\ref{itm:latency}. 
 Moreover, 
 the  \ac{pdf}  induced by $N\rightarrow \infty$ particles  does not necessarily converge to the true \ac{pdf}, and may depend on the sampling distribution ~\cite{moral1997nonlinear,del2000branching}. \acp{pf} thus require knowledge and the ability to approximate \eqref{eqn:StateEvolution} and \eqref{eqn:Measurements}. When either is mismatched, as pointed in \ref{itm:mismatch}, performance considerably degrades, possibly yielding an estimation bias that cannot be rectified solely through  increasing  $N$. 

Consequently, a main consideration when designing a \ac{pf} revolves around the choice of the importance sampling function $q(\myVec{x}^k)$, with the number of particles tuned to reach a desired  performance, and different \acp{pf} incorporate various methods to that aim. Such approaches, combined with dedicated resampling procedures, are the basis of the regularized \ac{pf} ~\cite{musso2001improving}, the \ac{apppf} ~\cite{ubeda2017adaptive}, the target resampling \ac{apppf} ~\cite{ubeda2017adaptive}, and the progressive proposal \ac{pf} ~\cite{bunch2013particle},  each with its own associated computational costs and accumulated latency.
\color{black}
%
The proliferation of \ac{pf} techniques, combined with availability of data in \ref{itm:settings1}-\ref{itm:settings2}, motivate exploring a complimentary data-aided approach to efficiently improve \ac{pdf} representation; one that can be integrated into any given \ac{pf} algorithm, as studied in the following section. 

\section{Learning Flock PF}
\label{sec:LF}
Here, we propose \ac{lf}, designed to enhance a given  \ac{pf} to meet \ref{itm:dist}-\ref{itm:NumTargets}, providing a solution that can be applied to almost any \ac{pf}. Accordingly, in our algorithm, for both the architecture and training, we utilize only the most basic elements that are common to \acp{pf}, the particles and weights. 
The high-level description of  \ac{lf}  is explained in Subsection~\ref{ssec:Rationale}. We elaborate on the architecture of the \ac{lf} module 
in Subsection~\ref{ssec:architecture}, and go through our novel loss function and  training procedures in Subsections~\ref{ssec:LossFunction}-\ref{ssec:Training}, respectively. We conclude with a discussion in Subsection~\ref{ssec:Discussion}.

\subsection{High-Level Description}
\label{ssec:Rationale} 
  We build on the understanding that   the key factors limiting the ability of \acp{pf} to adequately represent a \ac{pdf} when $N$ is limited (\ref{itm:latency}), and in the presence of stochasticity and model mismatches (\ref{itm:mismatch}), are encapsulated in the {\em complete} particles-weights set. 
We recognize that \ac{pf} algorithms typically handle each full trajectory $\myVec{x}_i^{0:k}$ separately~\cite{musso2001improving,ubeda2017adaptive,bunch2013particle,arulampalam2002tutorial,kronander2014robust}. Therefore, each particle evolves with only minimal consideration of other particles through the relativity of their weights (Steps~\ref{stp:for3}-\ref{stp:normalize} of Algorithm~\ref{alg:PF}). 
We thus propose to tackle the challenging factors mentioned in Subsection~\ref{ssec:Paticle} in a manner that can be integrated into existing \ac{pf} by a per iteration application of a correction term to each particle, which accounts for all particles. 
This methodology can prevent \acp{pf} typical deterioration of individual particles as time advances, aligning outliers and correcting inaccurate sampling, mitigating particle degeneracy and sample impoverishment. 
Doing so  leads to a substantial improvement of the capture of the state probability distribution along the trajectory, as we demonstrate in Section~\ref{sec:Experimental Study}.  

In particular, we use deep learning tools as correction terms via neural augmentation~\cite{shlezinger2020model,shlezinger2023model} {\em in conjunction with the pertinent \ac{pf} algorithm}, to modify a flock of particles by information sharing. 
 An augmentation example of our proposed \ac{lf} into a generic \ac{pf} is exemplified as Algorithm~\ref{alg:LF_aug_PF}. There,  $f_{\myVec{\theta}}(\cdot)$ represents the \ac{lf} \ac{dnn} with trainable parameters $\myVec{\theta}$. 
 This \ac{dnn} is trained to combine information of all particles in the flock  in order to correct each individual particle and overcome stochastic predicaments, as well as the need to accurately implement the importance sampling and weight adjustments. We next detail on the architecture of this \ac{lf} module and its training procedure.

\begin{algorithm}
    \caption{A generic \ac{lfpf} iteration on time $k$}
    \label{alg:LF_aug_PF} 
    \SetKwInOut{Input}{Input}  
    \Input{Particles-weights set $\{\myVec{x}_i^{k-1},w_i^{k-1}\}_{i=1}^{N}$, measurements $\myVec{z}^{k}$}  
    {
        \For{$i = 1, \ldots, N$}{
        \label{stp:LF_for1}
             Sample updated particles: \
             $\breve{\myVec{x}}_i^{k}\sim q\left(\myVec{x}^k|\cdot\right)$\;
              \label{stp:LF_advance}
        }
        \For{$i = 1, \ldots, N$}{
        \label{stp:LF_for2}
        Update weights: \
        $\breve{w}_i^{k}\leftarrow{{\pi\left(\breve{\myVec{x}}_i^{k}|\cdot\right)}/{q\left(\breve{\myVec{x}}_i^{k}|\cdot\right)}}$\label{stp:LF_reweight}\;
        }        
        \ac{lf} update of the particles-weights set:
        $\{\myVec{x}_i^{k},w_i^{k}\}_{i=1}^{N}\leftarrow \{\breve{\myVec{x}}_i^{k},\breve{w}_i^{k}\}_{i=1}^{N} + f_{\myVec{\theta}}(\{\breve{\myVec{x}}_i^{k},\breve{w}_i^{k}\}_{i=1}^{N})$\;
        \label{stp:LF_aug}
        \For{$i = 1, \ldots, N$}{
        \label{stp:LF_for3}
            Normalize: $w^k_i\leftarrow {w^k_i}/{\sum_{j=1}^{N}w^k_j}$ \; 
            \label{stp:LF_normalize}
        }
        \SetKwInOut{Output}{Output} 
        \label{stp:LF_output}
    \Output{Particles-weights set $\{\myVec{x}_i^{k},w_i^{k}\}_{i=1}^{N}$}
    }
\end{algorithm}

\begin{figure*}
\vspace*{-7mm}
    \includegraphics[width=\textwidth]{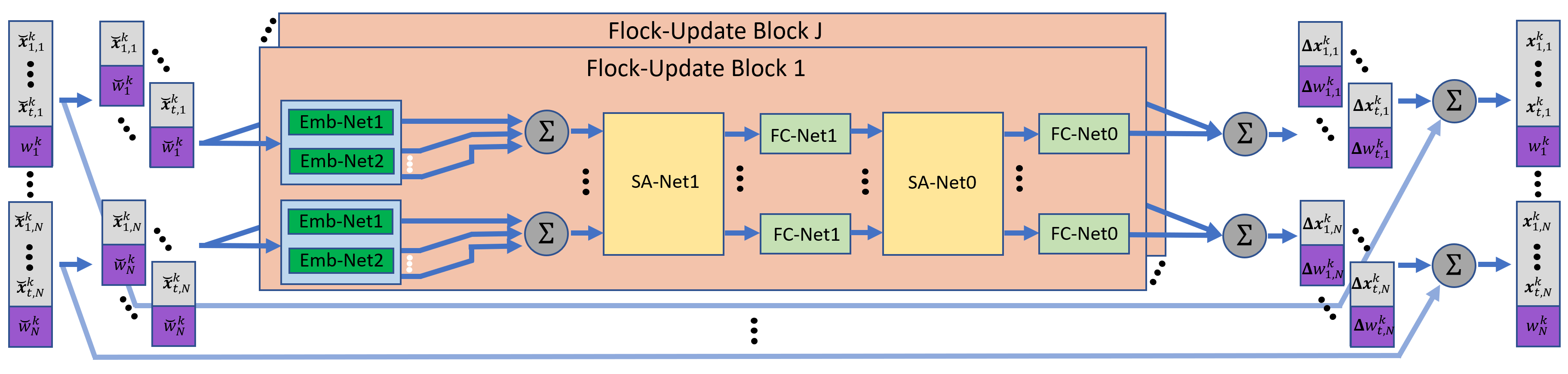}
    \caption{{\ac{lf} block architecture block diagram. A set of $N$ particle-weight pairs $\{\breve{\myVec{x}}_i^k, \breve{w_i}^k\}_{i=1}^N$ is decomposed into $N \times t$ sub-particles $\{\breve{\myVec{x}}_{j,i}^k,\breve{w_i}^k\}_{j,i=1}^{t,N}$, and processed by $J$ {\em flock-update} blocks in parallel.  Each block contains two parallel embedding networks in series with two \ac{sa} blocks and \acs{fc} layers, outputting a correction term to each full-particle. 
    }}
    \label{fig:DNN_architecture}
\end{figure*}

\subsection{Architecture}
\label{ssec:architecture} 
The \ac{lf} module, parameterized by $\myVec{\theta}$, is incorporated into a given \ac{pf} algorithm chosen for a specific task. 
As detailed in Algorithm~\ref{alg:LF_aug_PF} (Step~\ref{stp:LF_aug}), it maps a complete set of particles and weights $\{\breve{\myVec{x}}_i^k,\breve{w_i}^k\}_{i=1}^{N}$ into a {\em correction term}  $f_{\myVec{\theta}}(\{\breve{\myVec{x}}_i^{k},\breve{w}_i^{k}\}_{i=1}^{N})$. 
The \ac{lf} module design is based on the following considerations: 
$(i)$ every particle-weight correction should take into account all other particles and weights;
$(ii)$ the particles and weights pairs, and their update, is invariant to their ordering  and to the internal ordering of the sub-sates;  
$(iii)$ the \ac{lf} block should not induce considerable latency. 

Accordingly, we propose the architecture illustrated in Fig.~\ref{fig:DNN_architecture}. The architecture consists of  $J$ {\em flock-update} blocks, each combining two  parallel {\em sub-particle embedding} networks, with {\em permutation-invariant \ac{sa}} architecture. 
In the basic flock-update block, each state particle-weight pair  $[\breve{\myVec{x}}_i^k, \breve{w_i}^k] \in \mathbb{R}^{{d_{p}}+1}$ is separated to $t$ single sub-state sub-particles $\{[\breve{\myVec{x}}_{j,i}^k, \breve{w_i}^k]\}_{j=1}^t \in \mathbb{R}^{{d_{sp}}+1}$ (where the weight is shared among all sub-particles). These are transformed into two sub-particle embeddings of length $P\times 1$, {\em main} and {\em secondary}, by two separate embedding \ac{fc} networks, coined ${\rm Emb-Net1}$ and ${\rm Emb-Net2}$, respectively. 
For each of the $N$ particles, each sub-particle main embedding is added to the mean of the $t-1$ other sub-particles secondary embeddings to get its $t$ {\em full} embeddings of the same dimension $P\times 1$. 
Note that this procedure is invariant to the order of the sub-particles, thus maintaining the desired permutation invariance. 
Moreover, the same two embedding networks are used for all sub-particles, and thus the architecture is invariant to the number of sub-states $t$, satisfying~\ref{itm:NumTargets}.  

The full embeddings are then combined using dedicated \ac{sa} blocks (e.g., in the architecture illustrated in Fig.~\ref{fig:DNN_architecture}, two such blocks are used). In these blocks,  for  each of the $t$ sub-states, all $N$ full sub-particles embeddings from all particles are  updated in the same \ac{sa} layer (${\rm SA-Net}$) considering all $N$ full embeddings associated with the same sub-state. This layer is followed by an \ac{fc} network (${\rm FC-Net}$) applied in parallel to each of the $N \cdot t$ \ac{sa} outputs, thus maintaining the particle permutation invariance of the \ac{sa} layer. To get each particle update, at the output of the last \ac{fc} layer, the $t$ sub-states updates are concatenated while the weights are averaged. For
intermediate filtering stages applications where each particle's sub-state is assigned with its own weight, the initial weight sharing and final weight averaging can be skipped.

Inspired by multi-head attention mechanisms~\cite{vaswani2017attention}, we apply $J$ flock-update blocks in parallel. 
The $J$ outputs are summed up with the initial particle-weight pairs to get the final particle. 
 On flock-update block $J$, $J>1$, prior to the last \ac{fc} network, all embeddings for each sub-state are averaged, providing a single embedding per sub-state, that is turned into a single particle correction term at the final output of the block. 
That single particle term acts as baseline per sub-state and is added to all particles correction terms (from all other flock-update blocks), shifting their entire induced \ac{pdf} accordingly. 
 
We note that the architecture and its trainable parameters, as well as the described particles and sub-particles processing, is invariant to the number of sub-states $t$ and to their internal ordering (thus holding~\ref{itm:NumTargets}). They are also invariant to the number of particle-weight pairs $N$ and to their ordering, while enabling parallel operation (thus facilitating~\ref{itm:latency}).
Accordingly, the same trained architecture can be applied in settings with different numbers of sub-states and particles, e.g., an architecture trained for \ac{mtt} with a small number of targets (and/or particles) can be readily applied to track a larger number of targets (and/or particles), as we show in Section~\ref{sec:Experimental Study}.

    
    
\subsection{Loss Function}
\label{ssec:LossFunction} 

The training of  $f_{\myVec{\theta}}(\cdot)$ tunes $\myVec{\theta}$, encourages its correction term to satisfy two main properties: 
$(i)$ to gauge the accuracy of the state estimate \eqref{eqn:Estimate}; $(ii)$ to align the \ac{pdf} represented by  a set of  particles, and approximated using  \eqref{eqn:posterior}, with a desired \ac{pdf}.  
We next formulate the proposed loss assuming  supervised settings (\ref{itm:settings2}), after which we show how it can be specialized to unsupervised scenarios (\ref{itm:settings1}).

\subsubsection{Loss Formulation}
\label{sssec:LossForm}
The loss used for training $\myVec{\theta}$ based on data $\mySet{D}$ is comprised of two main parts, $\mathcal{L}^{\rm acc}(\cdot)$ and $\mathcal{L}^{\rm hm}(\cdot)$, corresponding to accuracy (Property $(i)$) and heatmap representation (Property $(ii)$), respectively. The resulting loss is given by
    \begin{align}
         \mathcal{L}_{\mySet{D}}(\myVec{\theta}) =  \frac{1}{|\mySet{D}|}\sum_{(\myVec{z}^{1:\kappa},\bar{\myVec{x}}^{1:\kappa}) \in \mySet{D} } \frac{1}{\kappa} \big( &\lambda_1\mathcal{L}^{\rm acc}(\myVec{z}^{ 1:\kappa},\bar{\myVec{x}}^{ 1:\kappa};\myVec{\theta}) \notag \\
         + &\lambda_2\mathcal{L}^{\rm hm}(\myVec{z}^{1:\kappa}{,\bar{\myVec{x}}^{1:k}};\myVec{\theta}) \big),
    \label{eqn:loss}
    \end{align}
    where $\lambda_1$ and $\lambda_2$ are non-negative hyperparameters balancing the contribution of each loss term. 

 Focusing on multiple sub-states tracking, for Property $(i)$, evaluated in $\mathcal{L}^{\rm acc}(\cdot)$, we employ the \ac{ospa} measure \cite{schuhmacher2008consistent}, often used to represent accuracy  in  \ac{mtt}. Specifically, we employ the \ac{ospa} distance  with order of $2$ and infinite cutoff, calculated per time-step, i.e., 
    \begin{equation}        
        \mathcal{L}^{\rm acc}(\myVec{z}^{ 1:\kappa},\bar{\myVec{x}}^{1:\kappa};\myVec{\theta}) =  \sum_{k=1}^{\kappa} {\rm OSPA}\left(\bar{\myVec{x}}^k, \hat{\myVec{x}}^k(\myVec{z}^{1:k}; \myVec{\theta})\right).
    \label{eqn:loss_acc}
    \end{equation}
In \eqref{eqn:loss_acc}, $\hat{\myVec{x}}^k(\myVec{z}^{1:k}; \myVec{\theta})$ represents the state estimation computed via \eqref{eqn:Estimate} and the corrected set of particles on time-step $k$. 

Property $(ii)$ is evaluated in $\mathcal{L}^{\rm hm}(\cdot)$, by comparing the distribution induced by the resulting particles with some oracle distribution. In particular, this loss term is comprised of the $\ell_2$ norms between the oracle true/desired posterior  and its  sub-states' per dimension  variances, denoted $\hat{p}_{\rm oracle}(\myVec{x}^k)$ and $\{\hat{V}_{{\rm oracle}|j}^k\}_{j=1}^t$, respectively,  and the corresponding reconstructed values obtained from the particles, corrected using the  \ac{lf} block. For the latter, the reconstructed \ac{pdf}, denoted $\hat{p}_{\myVec{\theta}}(\myVec{x}^k|\myVec{z}^{1:k})$, is computed via \eqref{eqn:posterior}, while the variances term of sub-particles $j$  denoted ${\rm Var}_j^k(\myVec{z}^{1:k}; \myVec{\theta})$ is computed as the $d_{sp}$ variances {across} the $d_{sp}$ dimensions of  the $N$ sub-particles of sub-state $j$. The resulting loss term is given by
    \begin{align}
        \mathcal{L}^{\rm hm}(\myVec{z}^{ 1:\kappa}&{,\bar{\myVec{x}}^{1:k}};\myVec{\theta}) =  \notag\\
         \sum_{k=1}^{\kappa}\Big( \| \hat{p}&_{\rm oracle}(\myVec{x}^k|\myVec{z}^{1:k} {,\bar{\myVec{x}}^k})-  \hat{p}_{\myVec{\theta}}(\myVec{x}^k|\myVec{z}^{1:k})\|_2 \notag\\
        &+\lambda_3 \sum_{j=1}^{t}{\|{\hat{V}_{{\rm oracle}|j}^k - {\rm Var}_j^k(\myVec{z}^{1:k}; \myVec{\theta})}\|_2}\Big),
    \label{eqn:loss_hm}
    \end{align}
where $\lambda_3$ is a hyperparameter. 
The \acp{pdf} comparison  is performed over a set of points in the state-space, hereby referred to as the {\em grid points}, selected in one of two ways that are detailed in Subsection~\ref{sssec:LossGrid}; and the variances comparison aims to align particles that stray too far out from the grid points region to be  accounted for.

\subsubsection{Evaluating the Loss}
\label{sssec:LossEval}
Evaluating \eqref{eqn:loss} requires, for each time instance $k$, access to the {true/}desired state $\bar{\myVec{x}}^{k}$, the oracle posterior  $\hat{p}_{\rm oracle}(\myVec{x}^k| \myVec{z}^{1:k} {,\bar{\myVec{x}}^k})$ and variances $\{\hat{V}_{{\rm oracle}|j}^k\}_{j=1}^t$. To obtain  the oracle posterior and variances, we use the pertinent \ac{pf} with $\tilde{N} \gg N$ particles, with the oracle variances attained as described for ${\rm Var}_j^k(\myVec{z}^{1:k}; \myVec{\theta})$ using the reference flow particles. Obtaining the oracle posterior and the true state differs based in the settings (\ref{itm:settings1} or \ref{itm:settings2}): 
\begin{enumerate}[label={O\arabic*}]
    \item \label{itm:opt1} {In the unsupervised settings \ref{itm:settings1}, we use the reference \ac{pf} with $\tilde{N}$ as a form of knowledge distillation. We utilize the reference flow particles and weights with a fixed Gaussian kernel,  $K_{j,i}\left(\cdot\right)\equiv K\left(\cdot\right)$, via  \eqref{eqn:posterior} to estimate $\hat{p}_{\rm oracle}(\myVec{x}^k)$, and via \eqref{eqn:Estimate}  for the desired state $\bar{\myVec{x}}^{}$.}
    \item \label{itm:opt2} In the supervised settings \ref{itm:settings2}, the   true state $\bar{\myVec{x}}^{}$ is available in the data, and $\hat{p}_{\rm oracle}(\myVec{x}^k)$  is isotropic multivariate Gaussian {distributions} centered in the true sub-states, with variances equal to the average ${\hat{V}_{{\rm oracle}|j}^k}$. 
\end{enumerate}

\begin{figure}
\vspace*{-5mm}
    \centering
        \includegraphics[width=1.0\columnwidth]{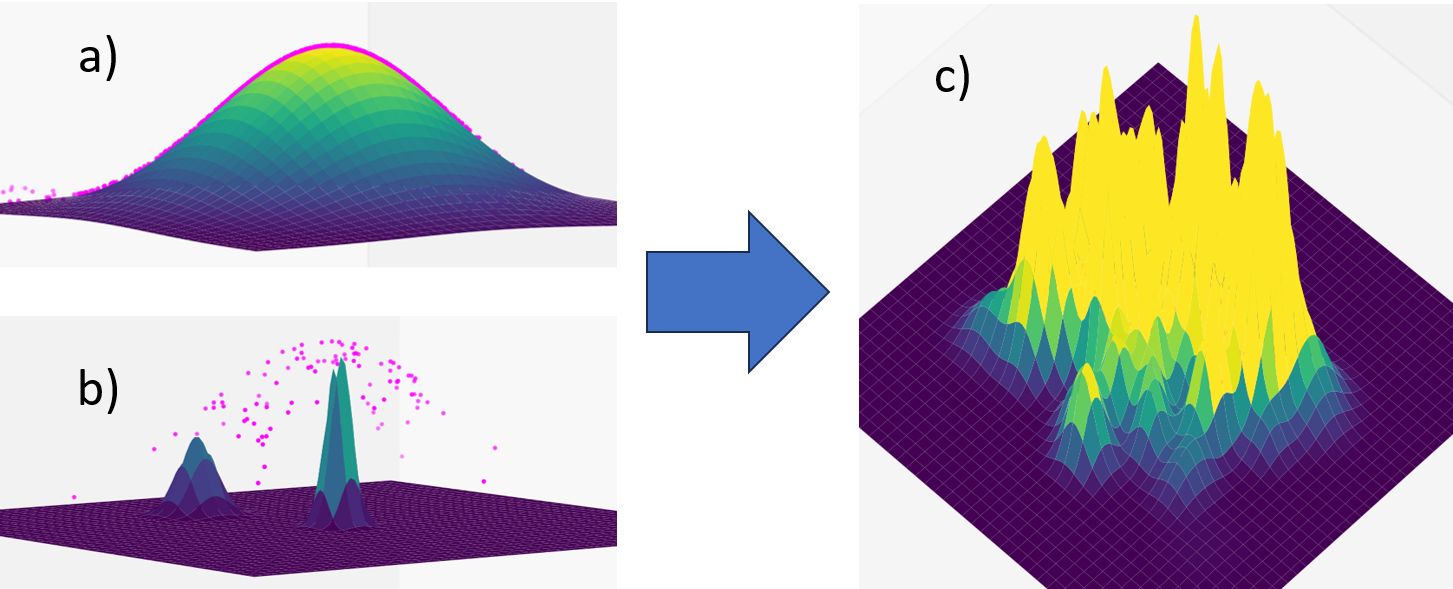}
    \caption{The construction of the actual \ac{pdf} $c)$ using the {\em adapting kernels} functions $b)$ based on the desired \ac{pdf} $a)$.}
    \label{fig:actual_pdf_reconstruction}
\end{figure}

The last remaining component needed to evaluate \eqref{eqn:loss} is the \ac{pdf} induced by the particles and their weights, $\hat{p}_{\myVec{\theta}}(\myVec{x}^k|\myVec{z}^{1:k})$. We do that using \eqref{eqn:posterior} and a set of $d_{sp}$-dimensional Gaussian kernel functions  $\{K_{j,i}(\cdot)\}_{j,i=1}^{t,N}$ we named {\em adapting kernels} illustrated in Fig.~\ref{fig:actual_pdf_reconstruction}. These Kernel functions are set to have covariances $\{\sigma_{j,i}^2\myMat{I}\}$, with  $\{\sigma_{j,i}\}$ determined such that all kernels have equal volume that sum up to $t$, and with peak height of ${p}_{\rm oracle}(\myVec{x}^k_{j,i})$. Using these settings, 
particles in high probability regions (where they are likely to have other particles close by)  will be assigned with high and narrow kernels, and  in case that they are too close, will be encouraged to disperse by the locally lower desired \ac{pdf} on the loss. 
Similarly, particles in low probability regions (where they are likely to be isolated) will be assigned with wide  kernel functions, and so, will be encouraged by the loss to be more evenly distributed. This formulation supports adapting the grid points resolutions according to the particles predicted density, as well as boosts a smooth and accurate \ac{pdf} reconstruction.

\subsubsection{PDF Comparison}
\label{sssec:LossGrid}
\begin{figure}
\vspace*{-5mm}
    \centering
        \includegraphics[width=1.0\columnwidth]{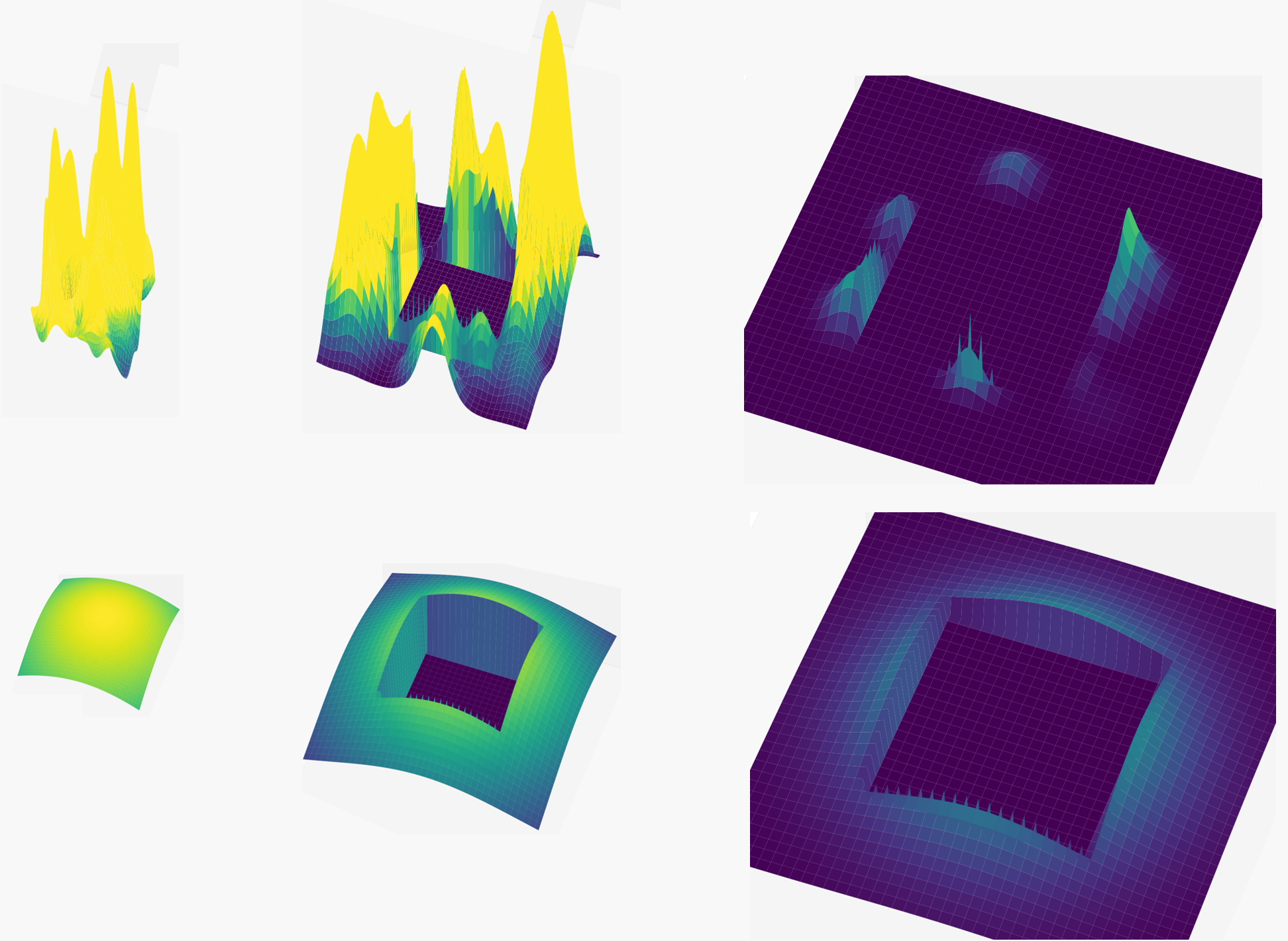}
    \caption{{\em staged meshgrid}: The heatmap loss is calculated as the sum of  $L$ squared error between $L$ pairs of heatmaps grids, desired (top row) and actual (bottom row), on different resolutions and scales.}
    \label{fig:staged_heatmap}
\end{figure}

The comparison of the two \acp{pdf}, i.e., the term $\| \hat{p}_{\rm oracle}(\myVec{x}^k|\myVec{z}^{1:k}{,\bar{\myVec{x}}^k}) - \hat{p}_{\myVec{\theta}}(\myVec{x}^k|\myVec{z}^{1:k})\|_2$ in \eqref{eqn:loss_hm}, is approximated by computing their differences over a grid, and the selection of the grid points highly impacts the usefulness of the loss.  
We present two methods for choosing the {\em grid~point}s around the sub-states locations:
 
{\bf Staged Meshgrid:} 
The first method selects the grid points to be evenly distributed on a $d_{sp}$-dimensional cube, or a meshgrid, while adaptively tuning its scale, resolution and location. As illustrated in Fig.~\ref{fig:staged_heatmap}, the \acp{pdf} in \eqref{eqn:loss_hm}  are compared $L>0$ times with $L$ different grids of the same size and different resolutions, co-centered according to $\bar{\myVec{x}}_j^k$ and $\hat{\myVec{x}}_j^k(\myVec{z}^{1:k}; \myVec{\theta})$. The overlaps between the different resolutions is accounted for once on the higher resolution, that is also considered in the grid points weighting in the loss.
 
Using {\em adapting kernels} together with {\em staged meshgrid} tackles  two of the main challenges encountered  in \acp{pf} -- sample impoverishment and particle degeneracy -- as well as stochastic outliers.  
Sample impoverishment tends to happen close to the center of the support of the \ac{pdf}, or in high resolution meshgrid stages, that can accommodate the more narrow kernels that are likely to be used there. 
Particle degeneracy usually occurs in the edges of the support of the \ac{pdf} where the covered area is bigger and probability density is lower; there,  the 
adapting kernels will tend to be wider and cover larger area so can be captured by the lower resolution stages. We implement and  test the staged meshgrid method  in Subsection~\ref{ssec:RadarTargetTracking}.

{\bf Random Grid:}  
Despite the aforementioned benefits of our  staged meshgrid approach, it may be computationally intensive in some settings, particularly when the sub-state is of a high dimension $d_{sp}$. 
An alternative approach uses a {\em random grid}. 
Here, instead of the evenly spread points as on a meshgrid, we randomize sampling points according to a predetermined distribution. 
Similar to the staged meshgrid, the random grid points sampling distribution can have different density  in different regions to accommodate changing resolutions and tackle particle degeneracy and sample impoverishment. 
Pseudo-random sampling points will result in more evenly distributed points and a better cover of the area of interest, and may induce  faster and better learning, however this is left for future work.  The random grid method is numerically evaluated  in Subsection~\ref{ssec:StateEstimation}.

\subsection{Training}
\label{ssec:Training}

We note that the loss function in \eqref{eqn:loss} expresses a sum of $\kappa$ time-steps components of \eqref{eqn:loss_acc}, and \eqref{eqn:loss_hm},  computed separately at each time-step $k$ when an \ac{lf} block with parameters $\myVec{\theta}$ is applied. We can leverage this property to enable training \ac{lf} on each (or selected) time-step seperately using conventional \ac{sgd}-based learning, while avoiding the need for differentiability between iterations (and avoiding the need to backpropagate through sampling operators using special sampling adaptations ~\cite{scibior2021differentiable,zhu2020towards}, or by introducing approximated operators~\cite{jonschkowski2018differentiable,chen2021differentiable}).

A candidate training method based on mini-batch \ac{sgd} for an  \ac{lfpf} is formulated in Algorithm \ref{alg:Train}. There, the \ac{lfpf} is executed in parallel with the reference \ac{pf} flow (Steps~\ref{stp:Train_aug}, \ref{stp:Train_referance}), that maintain particles-weights sets for each trajectory, respectively denoted $\mySet{P}_q^k, \mySet{R}_q^k$  for the $k$th time step of the $q$th batch. Training is done by loss calculation (Steps~\ref{stp:Train_oracle}-\ref{stp:Train_loss}) and its backpropagation (Step~\ref{stp:Train_update}), while nullifying gradient propagation between iterations and through sampling operations. To achieve a fast and continuous convergence, on each time-step  $k$ we mix the loss of  different trajectories  by running a bigger batch of trajectories $\mathcal{D}_q$, while training on a random subset of the batch that changes between time-steps (Step \ref{stp:Train_rand_subset}). For tracking stabilization during training, particularly at the beginning, once an estimated sub-state strays farther than a predetermined distance from the desired sub-state (Step \ref{stp:Train_condition}) the whole trajectory is eliminated from $\mathcal{D}_q$ for the training on the following time-steps (Step \ref{stp:Train_eliminate}). 
\color{black}

\begin{algorithm}
    \caption{\ac{lfpf} training}
    \label{alg:Train} 
    \SetKwInOut{Initialization}{Init}
    \Initialization{Set augmented and reference \acp{pf};\\
    Initialize \ac{lf} parameters $\myVec{\theta}$;\\
    Fix learning rate $\mu>0$; \\
    Set sub-state stray distance threshold $\zeta$;}
    \SetKwInOut{Input}{Input}  
    \Input{Training set  $\mathcal{D}$, initial states $\{\bar{\myVec{x}}^0\}$}  
    \For{$e = 1, \ldots$ \textbf{ until convergence}}{
        Randomly divide $\mathcal{D}$ into $Q$ batches $\{\mathcal{D}_q\}_{q=1}^Q$\;
        \For{$q = 1, \ldots, Q$}{  
            {Initialize \ac{lfpf} and reference \ac{pf} $|\mathcal{D}_q|$  particles-weights sets $\mathcal{P}^0_q$ and $\mathcal{R}^0_q$  to $\{\bar{\myVec{x}}^0\}$;}\\
            \label{stp:Train_zero_loss}
            \For{$k = 1, \ldots, \kappa$}{
                {Apply~\ac{lfpf}~with~$\myVec{\theta}$~to~$\mathcal{D}_q$,~$\mathcal{P}^{k-1}_q$~to~get~$\mathcal{P}^k_q$}\; 
                \label{stp:Train_aug}
                {Apply reference \ac{pf} to $\mathcal{D}_q$, $\mathcal{R}^{k-1}_q$ to get $\mathcal{R}^{k}_q$}\; 
                \label{stp:Train_referance}
                For a random sub-batch $\tilde{\mathcal{D}}_{q}^k \subseteq \mathcal{D}_q$:
                \label{stp:Train_rand_subset}\\                
                \Indp 
                Evaluate $\hat{V}_{{\rm oracle}}^k$ and $\hat{p}_{\rm oracle}(\myVec{x}^k)$ using $\mathcal{R}^k_q$ and \ref{itm:opt1} or \ref{itm:opt2}\;
                \label{stp:Train_oracle} 

                Evaluate ${\rm Var}^k(\myVec{z}^{1:k}; \myVec{\theta})$ and $\hat{p}_{\myVec{\theta}}(\myVec{x}^k|\myVec{z}^{1:k})$ using $\mathcal{P}^k_q$ and $\hat{p}_{\rm oracle}(\myVec{x}^k)$\;
                \label{stp:Train_actual} 
                
                Calculate loss $\mathcal{L}_{{\tilde{\mathcal{D}}_{q}^k}}(\boldsymbol{\theta})$ via \eqref{eqn:loss}\;
                \label{stp:Train_loss}                    
                Update  $\boldsymbol{\theta}\leftarrow \boldsymbol{\theta} - \mu\nabla_{\boldsymbol{\theta}}\mathcal{L}_{{\tilde{\mathcal{D}}_{q}^k}}(\boldsymbol{\theta})$\;    
                \label{stp:Train_update}
                \Indm 
                \For{{\rm each} $(\myVec{z}^{1:\kappa}, \bar{\myVec{x}}^{1:\kappa})\in \mySet{D}_q$}{
                    \If {$\exists j$ {\rm s.t.} $\|\bar{\myVec{x}}_j^k, \hat{\myVec{x}}_j^k(\myVec{z}^{1:k}; \myVec{\theta})\|\geq \zeta$
                    \label{stp:Train_condition}
                    }
                        {
                        Remove $\mathcal{D}_q \leftarrow \mathcal{D}_q / \{(\myVec{z}^{1:\kappa}, \bar{\myVec{x}}^{1:\kappa})\}$\; 
                        \label{stp:Train_eliminate}              }
                }
            }
        }
    }
    \KwRet{$\boldsymbol{\theta}$}
\end{algorithm}

\subsection{Discussion}
\label{ssec:Discussion} 

The proposed trainable \ac{lf} module is trained as part of a specific \ac{pf}, and possibly with an additional more complex \ac{pf} serving as reference. 
Nonetheless, once trained, it can be readily transferred into alternative \acp{pf} that meet \ref{itm:dist}, as demonstrated in Subsection~\ref{ssec:StateEstimation}. 
Moreover, by harnessing supervised data, the \ac{lfpf} can also cope with \ref{itm:mismatch} without  incorporating additional mechanisms to explicitly mitigate it as in ~\cite{uney2015cooperative}; while enabling the pertinent \ac{pf} to operate with a smaller number of particles with minimal computational cost to meet \ref{itm:latency}.  
As discussed in Subsection~\ref{ssec:architecture}, the \ac{lf} architecture is invariant to the number of sub-states $t$, and, e.g., the same module can be trained and evaluated for tracking a different number of sub-states, thus meeting \ref{itm:NumTargets}. 
These properties are consistently demonstrated in Subsection~\ref{ssec:RadarTargetTracking}.

Our \ac{lf} module, trained with a specific number of particles, allows \acp{pf} to operate reliably with different, and even lower values of $N$, boosting low latency filtering. 
However, the application of \ac{lf} comes at the cost of some excessive complexity, depending on the parameters $J$, $d_{sp}$, $t$, $N$, $P$, as well as the {\em flock-update} parameters, number of sub-embeddings $E$ ($1$, or $2$ to include secondary embedding), number of \ac{sa} blocks $S$, and \ac{fc} width multiplier $B$. 
The total \ac{fpm} requirement per particle can be divided to three parts with  accordance to Fig.~\ref{fig:DNN_architecture}: broadly speaking, $(i)$ the embedding networks ${\rm Emb-Net1/2}$ require ${J  E t\cdot  B  P \left(P+d_{sp} +3 B P\right)}$ \ac{fpm}; $(ii)$ the \ac{sa} networks ${\rm SA-Net0/1}$ involve $ J  S t \cdot 2 P^2  \left(2 +N/P+B +B^2\right)$; and $(iii)$  the final \ac{fc} layers require $t \cdot B P\left( B P +d_{sp}\right)$ \ac{fpm}.  
While this excessive complexity varies considerably with the system parameters, it is noted that the operation of such compact \acp{dnn} is highly suitable for parallelization and hardware acceleration, that often notably reduces inference speed of \acp{pf} with similar performance, as demonstrated in Section~\ref{sec:Experimental Study}.  

Our proposed \ac{lf} jointly corrects particles and their weights utilizing only the particles and weights (that are merely updated by the \ac{lf} block). 
This boosts versatility and flexibility enabling a trained \ac{lf} block to be conveniently integrated and apply a fix to a flock at any point on different types of \acp{pf} flow, whether classical or \ac{dnn} augmented, or even to a different \ac{pf} than the one it was trained on (as shown in Subsection~\ref{ssec:StateEstimation}). 
Nonetheless, one can potentially extend \ac{lf} to process additional information when updating the particles, such  as past particles (smoothing), the measurements, and data structures~\cite{buchnik2024gsp}. 
Moreover, the \ac{lf} block can  be adapted to change its output, for instance, to include an estimated state  or additional embedding for other purposes. 
It  can possibly also be integrated in  \acp{pf} involving detection (e.g., track-before-detect~\cite{ubeda2017adaptive}). 
We leave these extensions for future work.

\section{Experimental Study}
\label{sec:Experimental Study}
In this section, we numerically evaluate our \ac{lf} algorithm in terms of performance and latency. We first  compare it to  state-of-the-art \ac{pf} neural augmentation for synthetic state estimation  in Subsection~\ref{ssec:StateEstimation}. Then, in Subsection~\ref{ssec:RadarTargetTracking} we combine our \ac{lf} with a well-established \ac{apf} for a radar \ac{mtt} setup\footnote{The source code and hyperparameters used in this experimental study is available  at \url{https://github.com/itainuri/LF-PF_MTT.git}.}. All timing/latency tests were computed on the same processor, AMD EPYC 7343 16-Core 3.2GHz.

\subsection{Synthetic State Estimation}
\label{ssec:StateEstimation} 
Here, we evaluate \ac{lfpf} in comparison to state-of-the-art neural augmentation of \acp{pf}, based on algorithm unrolling ~\cite{gama2022unrolling}. Our goal here is to showcase the performance gains of our proposed \ac{lf} algorithm compared to alternative types of \ac{dnn} augmentations, and that its operation is complementary to other enhancements of \acp{pf}. 

\subsubsection{Simulation Setup}
We adopt the synthetic simulation setup used in~\cite{gama2022unrolling}. 
Here, the state is comprised of $d_{sp}=10$ dimensional variables, and is tracked based on observations  $\myVec{z}^k$  of dimension $d_m=8$ over $\kappa=12$ time-steps. 
The state evolution model is given by 
    \begin{equation}        
            \myVec{x}^k=\phi(\myMat{A}\myVec{x}^{k-1})+\myVec{v}^k,
    \label{eqn:UrMotionModel}
    \end{equation} 
    while the observation model is
    \begin{equation}        
            \myVec{z}^k=\myMat{C}\myVec{x}^{k}+\myVec{e}^k.       
    \label{eqn:UrSensorModel}  
    \end{equation}
    In \eqref{eqn:UrMotionModel}-\eqref{eqn:UrSensorModel}, $\myMat{A}\in\mathbb{R}^{d_{sp}\times d_{sp}}$ and $\myMat{C}\in\mathbb{R}^{d_m\times d_{sp}}$ are fixed matrices taken from \cite[Ch. 4.1]{gama2022unrolling} and the noise vectors $\myVec{v}^k$ and $\myVec{e}^k$ are zero-mean with covariances $\myMat{\Sigma}_v$ and $\myMat{\Sigma}_e$, respectively. 
    Specifically, the noise distributions and the mapping $\phi(\cdot)$ are taken from the following setups:
\begin{enumerate}[label={X\arabic*}]
    \item \label{itm:LinGauss} A linear Gaussian system, where $\phi(\myVec{x})=\myVec{x}$ and $v^k$ and $w^k$ are zero-mean white Gaussian noises with fixed covariance matrices  $\myMat{\Sigma}_v$ and $\myMat{\Sigma}_e$ taken from \cite[Ch. 4.1]{gama2022unrolling}.
    \item \label{itm:NonLinGauss} Gaussian noise  as in \ref{itm:LinGauss}, with non-linear $\phi(\cdot)$ given by the absolute value function (taken element-wise).
    \item \label{itm:UniformNoise} A non-Gaussian system, where  $\phi(\myVec{x})=\myVec{x}$, with non-Gaussian uniform noise and mismatched assumed covariance matrices $\myMat{\Sigma}_v=\sigma^2\myMat{I}_{d_{sp}}$ and $\myMat{\Sigma}_e=\sigma^2\myMat{I}_{d_m}$.
\end{enumerate}
The noise covariances have the same $\ell_2$ norm, i.e., $\|\myMat{\Sigma}_v\|_2 = \|\myMat{\Sigma}_e\|_2$, dictating the \ac{snr}. 

\subsubsection{Estimation Algorithms}
Our main baseline \ac{pf} throughout this section is the \ac{sispf}~\cite{doucet2000sequential}. This estimator implements Algorithm~\ref{alg:PF} with a Gaussian sampling distribution $q(\myVec{x}^k)$, whose inverse covariance is $\myMat{\Sigma}^{-1}=\myMat{\Sigma}_v^{-1} +\myMat{C}^T\myMat{\Sigma}_e^{-1}\myMat{C}$, and mean is $\myVec{\mu}=\myMat{\Sigma}\left(\myMat{\Sigma}_v^{-1}\phi\left(\myMat{A}\myVec{x}_i^{k-1}\right)+\myMat{C}^T\myMat{\Sigma}_e^{-1}\myVec{z}^k\right)$.

Based on the formulation of the \ac{sispf} to the above models, we compare the following estimation algorithms:
\begin{itemize}
    \item \ac{sispf}  with $N=25$ and with $N=300$ particles.
    \item An \ac{lfsispf} with $N=25$ particles.  With $d_{sp}$  dictated by the settings, and with $P=64$, $B=2$, $J=2$, and $S=2$, the  \ac{lf} module has  $4.19\cdot{10}^5$ trainable parameters $\myVec{\theta}$, and $4.23\cdot{10}^5$ \ac{fpm} per particle per iteration with $25$ particles. The reference \ac{pf} used for training an is \ac{sispf} with $\tilde{N}=300$ particles, with the heatmap loss \acp{pdf} in \eqref{eqn:loss_hm} compared over $4\cdot 10^6$ {\em random grid} points with Gaussian distribution. 
    \item The \ac{urpf}  of \cite{gama2022unrolling}, which unrolls $\kappa$ \ac{pf} time-steps. All steps use the same covariance \ac{dnn} and each step uses a  dedicated mean \ac{dnn} for a multivariate normal importance sampling function.  
    The \acp{dnn} receive a particle and the measurements as input, and have $1.41\cdot (1+\kappa) \cdot {10}^5$ learnable parameters $\myVec{\xi}$, and require  $2.82\cdot{10}^5$ \ac{fpm} per particle per $25$ particles iteration.
    \item \ac{lfurpf}, Augmenting the trained data-based \ac{urpf} with an \ac{lf} module, trained with the model-based \ac{sispf} (both with the same $\myVec{\theta}$ and  $\myVec{\xi}$).
\end{itemize}All data-driven estimators were trained in an unsupervised manner: \ac{lfsispf} via \ref{itm:opt1}, and \ac{urpf} as described in~\cite{gama2022unrolling}. We use the \ac{mse} as our performance measure, evaluated over a $100$ test trajectories, a $100$ times each. 
For each scenario we trained the \ac{lfsispf} model on $\kappa=15$ long trajectories, {keeping the best weights according to time-step $k=12$, and} considering time-steps $k=9-15$ for the loss. 
For each setup, training  was initially done with \ac{snr}=0~dB and fine-tuned for each \ac{snr}.
While the same \ac{lf} pre-trained weights were applied on all trajectories of the same experiment, the \ac{urpf} was trained per test trajectory (per experiment). Specifically, we trained \ac{urpf}    for $200$ epochs on that $\kappa=12$ time-steps trajectory without resampling, keeping the best weights for testing according to two criteria, $\myVec{\xi}_1$ for the full trajectory average, and $\myVec{\xi}_2$ for the last time-step average.  
\color{black}

\subsubsection{Results} 
We test each of the benchmarks in all three settings with different \acp{snr}. All benchmarks have particles resampling procedure cooperated to their iteration flow, and testing is performed twice, once with $N_{\rm th}=N//3$ and once with $N_{\rm th}=0$ (without resampling). For clarity considerations we chose to present the experiment with the better results out of the two, where the \ac{sispf} with $N=300$ particles was always implemented with resampling.


\begin{figure}
\centering
\begin{subfigure}[t]{.49\columnwidth}
    \centering
    \includegraphics[width=\textwidth]{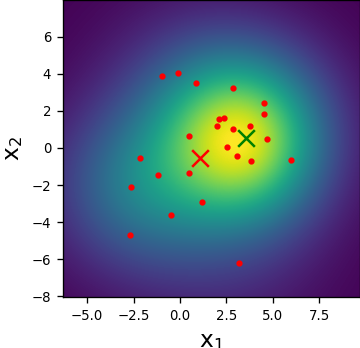}  
    \caption[]{Particles and \ac{pdf}, \ac{lf} input}
    \label{fig:heatmap_rand_grid_before_nn3}
\end{subfigure}
\hfill
\begin{subfigure}[t]{.49\columnwidth}
    \centering
    \includegraphics[width=\textwidth]{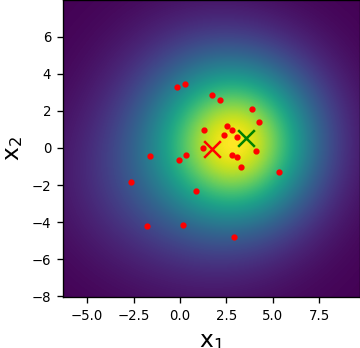}  
    \caption[]{Particles and \ac{pdf}, \ac{lf} output}
        \vspace*{-2mm}
    \label{fig:heatmap_rand_grid_after_nn3}
\end{subfigure}
\vspace*{+5mm}

\begin{subfigure}[t]{.49\columnwidth}
    \centering
    \includegraphics[width=\textwidth]{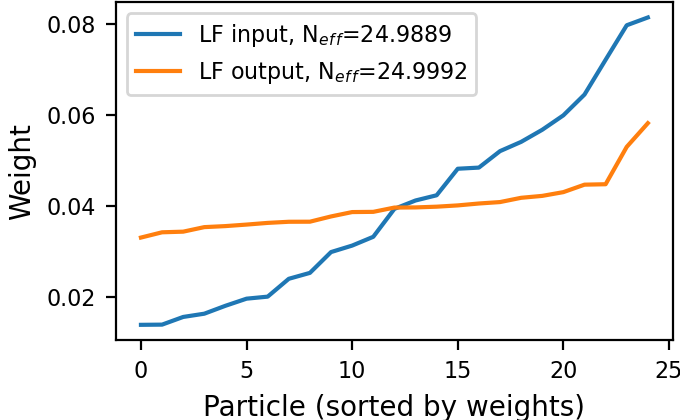}  
    \caption[]{Sorted weights.}
    \label{fig:sorted_wts_rand_grid}
\end{subfigure}
\begin{subfigure}[t]{.49\columnwidth}
    \centering
    \includegraphics[width=\textwidth]{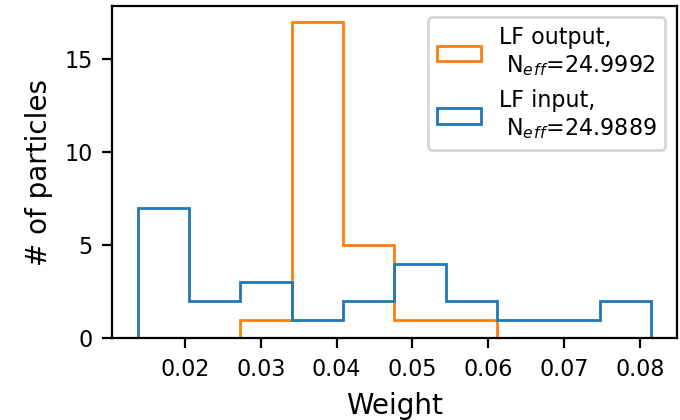}  
    \caption[]{Weights histograms.}
    \label{fig:wts_hists_rand_grid}
\end{subfigure}
\vspace*{+4mm}
\caption{\ac{lf} particles and weights adjustment example with the \ac{lfsispf} in the \ref{itm:UniformNoise} settings of Subsection~\ref{ssec:StateEstimation}.  Top: reconstructed \ac{pdf} cross-section and particles at the input (Fig.~\ref{fig:heatmap_rand_grid_before_nn3}) and output (Fig.~\ref{fig:heatmap_rand_grid_after_nn3}) of the \ac{lf} block. The particles projection on $[x_1,x_2]$ plane are marked with red dots, and the desired state's and its estimate's projections are marked with green and red crosses, respectively.
Bottom: the particles' sorted weights (Fig.~\ref{fig:sorted_wts_rand_grid}) and weights histograms (Fig.~\ref{fig:wts_hists_rand_grid}) at the input and output of the \ac{lf} module.}
\label{fig:rand_grid_parts_update}
\end{figure}

We first visualize the ability of the \ac{lf} module in forming desirable particle patterns. To that aim, we observe the operation of a trained \ac{lf} module for the settings \ref{itm:UniformNoise} at a given time instance $k=7$ and \ac{snr} of $0$ dB.  We illustrate in Figs.~\ref{fig:heatmap_rand_grid_before_nn3}-\ref{fig:heatmap_rand_grid_after_nn3}  how the trained \ac{lf} block herds the particles closer to the desired location (on  $x_1$ and $x_2$ axes), and how the new induced \ac{pdf} is more confined and the estimation is more accurate. 
We also examine the weights at the input and output of our \ac{lf} block in  Figs. \ref{fig:sorted_wts_rand_grid}-\ref{fig:wts_hists_rand_grid}. There, we observe a reduction in the variance of the weights and an increase in the effective number of particles  $N_{\rm eff}$ (without explicitly encouraging it via the loss), which potentially indicate an improved capture of the \ac{pdf}. This constant correction of the particles and weights alleviates the degeneracy phenomenon, as can be seen by the initial high $N_{\rm eff}$ on time-step $k=7$.

\begin{figure*}
\centering
\vspace*{-6mm}
\begin{subfigure}{.32\textwidth}
    \centering
    \includegraphics[width=1.0\linewidth]{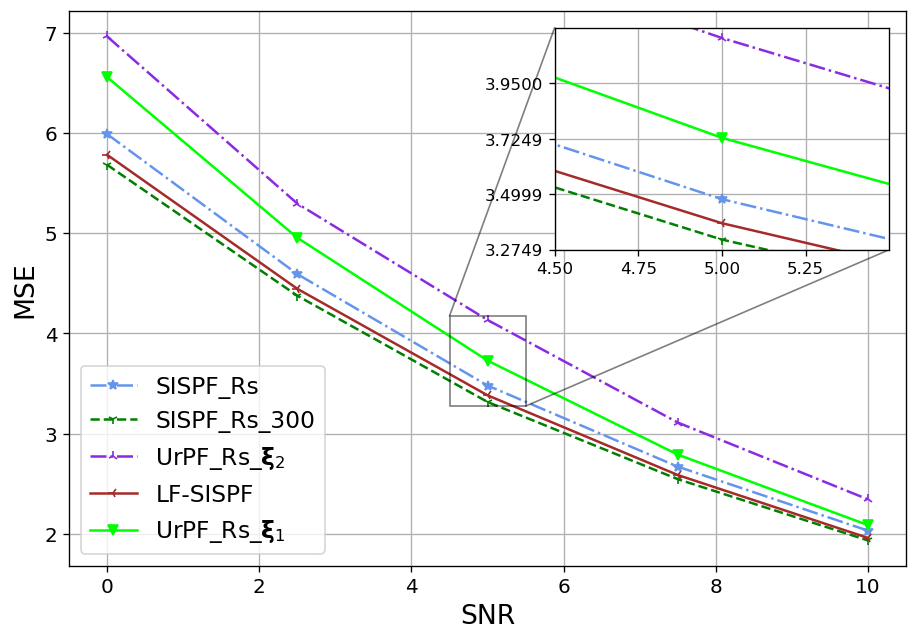}  
    \caption{\ref{itm:LinGauss}, entire trajectory accuracy}
    \label{sfig:UR_avg}
\end{subfigure}
\begin{subfigure}{.32\textwidth}
    \centering
    \includegraphics[width=1.0\linewidth]{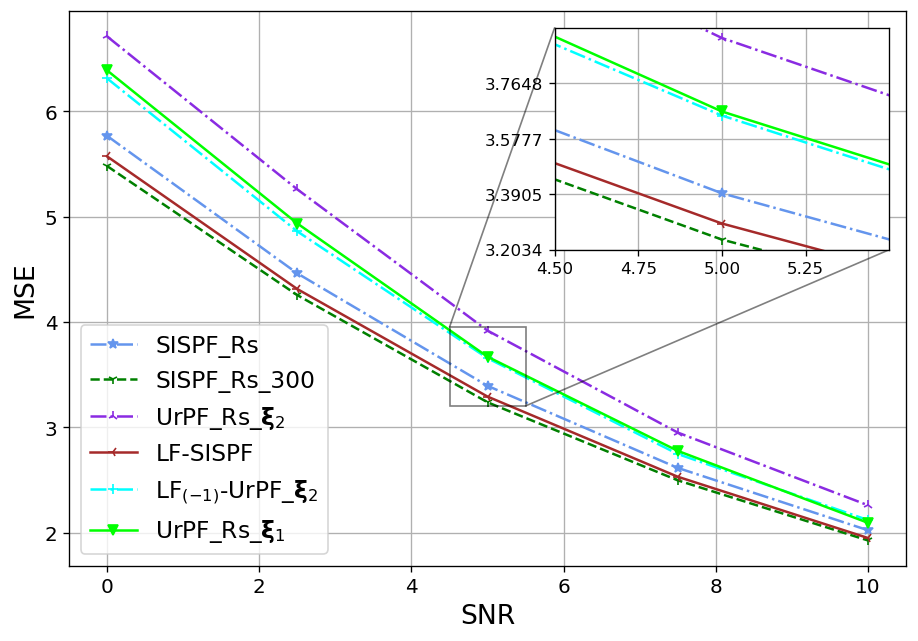}  
    \caption{\ref{itm:NonLinGauss},  entire trajectory accuracy}
    \label{sfig:UR_avg_nonlin}
\end{subfigure}
\begin{subfigure}{.32\textwidth}
    \centering
    \includegraphics[width=1.0\linewidth]{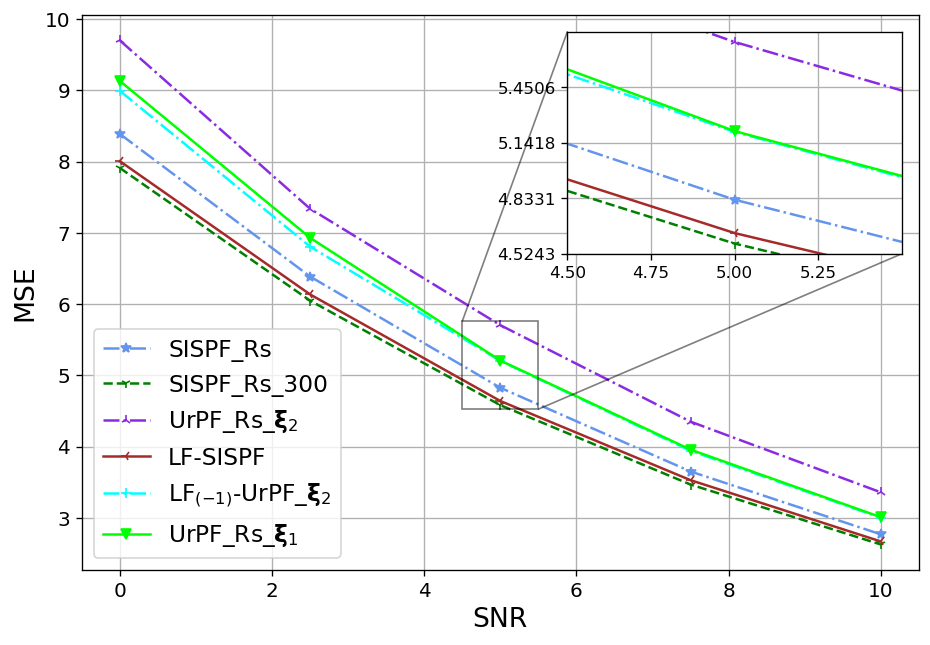}  
    \caption{\ref{itm:UniformNoise},  entire trajectory accuracy}
    \label{sfig:UR_avg_nongauss}
\end{subfigure}
\vspace*{+5mm}


\begin{subfigure}{.32\textwidth}
    \centering
    \includegraphics[width=1.0\linewidth]{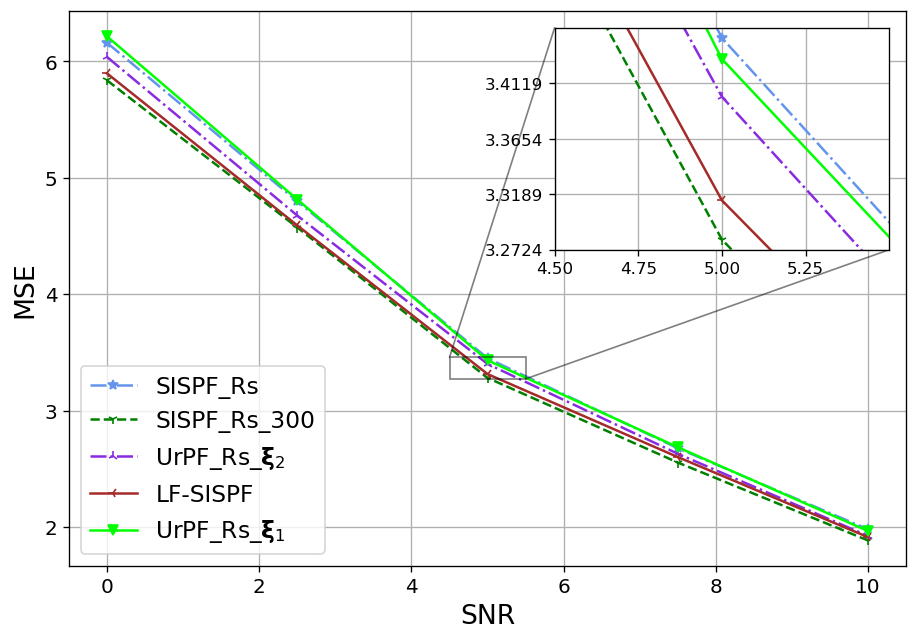}  
    \caption{\ref{itm:LinGauss}, last time-step accuracy}
    \label{sfig:UR_last}
\end{subfigure}
\begin{subfigure}{.32\textwidth}
    \centering
    \includegraphics[width=1.0\linewidth]{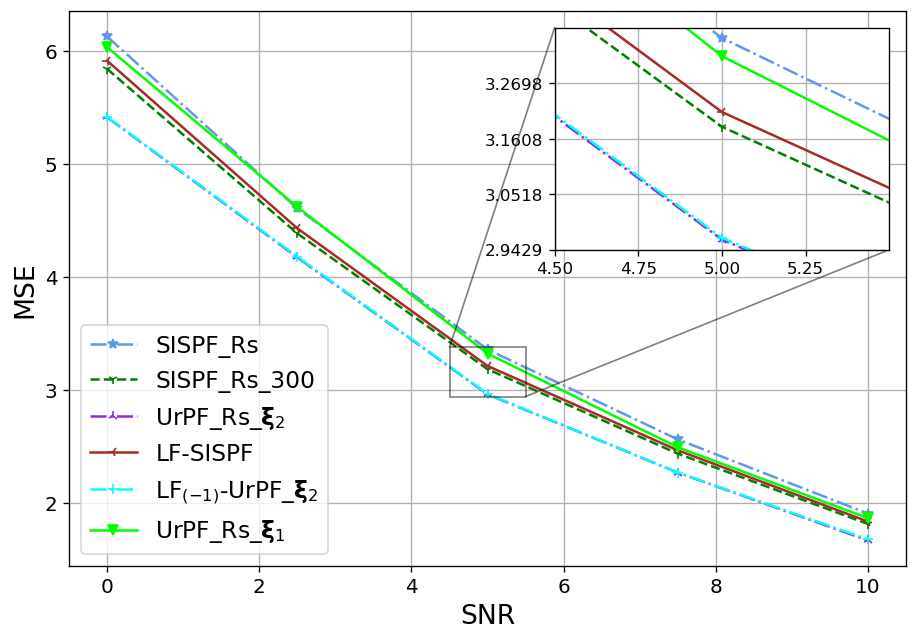}  
    \caption{\ref{itm:NonLinGauss}, last time-step accuracy}
    \label{sfig:UR_last_nonlin}
\end{subfigure}
\begin{subfigure}{.32\textwidth}
    \centering
    \includegraphics[width=1.0\linewidth]{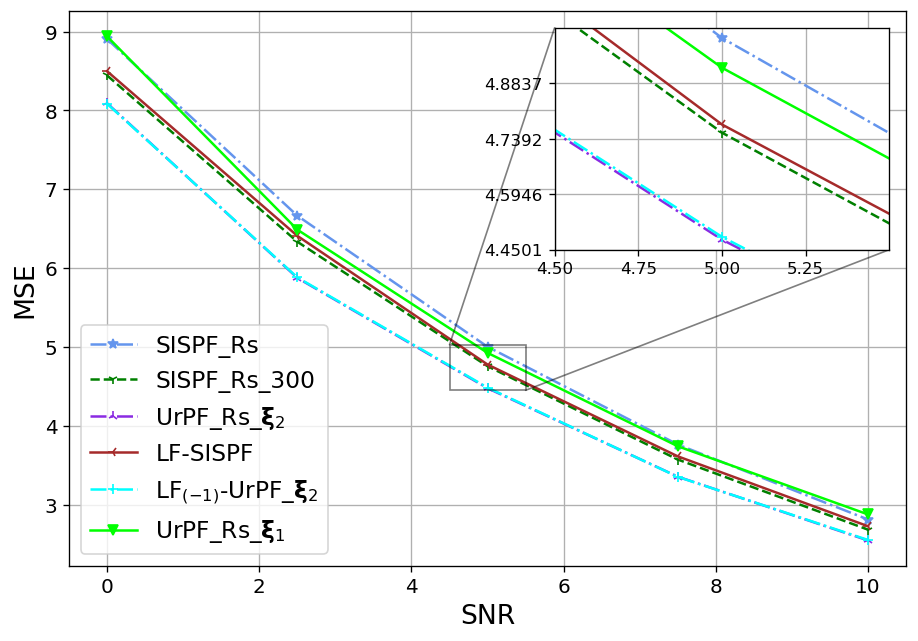}  
    \caption{\ref{itm:UniformNoise}, last time-step accuracy}
    \label{sfig:UR_last_nongauss}
\end{subfigure}
\vspace*{+5mm} 
\caption{Overall (top) and last time-step $k=\kappa$ (bottom) tracking accuracy for settings \ref{itm:LinGauss}-\ref{itm:UniformNoise} over $\kappa=12$ time-steps trajectories.  \ac{urpf}  is utilized for overall and last time-step accuracy with its respective task oriented trained parameters $\myVec{\xi}_1$  or $\myVec{\xi}_2$. ($\_Rs$) implies resampling with $N_{\rm th}=N//3$, and ($LF-$) or ($LF_{(-1)}-$) implies the utilization of the \ac{lf} including or excluding the last time-step. $SISPF\_Rs\_300$ was tested with $N=300$ particles, while all other benchmarks were tested with $N=25$.}
\label{fig:LF_vs_Unrolling}
\end{figure*}

We next show that the favourable particles and weights alternations achieved by \ac{lf} on a single time-step, as presented in Fig.~\ref{fig:rand_grid_parts_update}, translate into improved tracking accuracy. To that aim, we report in the  Fig. \ref{fig:LF_vs_Unrolling} the resulting \ac{ospa} values achieved by the considered tracking algorithms for settings \ref{itm:LinGauss}-\ref{itm:UniformNoise} in recovering the entire trajectory (Figs.~\ref{sfig:UR_avg}-\ref{sfig:UR_avg_nongauss}) and in recovering solely the final time-step $k=12$ (Figs.~\ref{sfig:UR_last}-\ref{sfig:UR_last_nongauss}), being the task for which \ac{urpf} is originally designed for, in \cite{gama2022unrolling}.

As observed in Fig.~\ref{fig:LF_vs_Unrolling}, on all experiments the \ac{lfsispf} notably improves the \ac{sispf} and achieves results that are close to those of its teacher \ac{pf}, namely, the  \ac{sispf} with $N=300$. In the straightforward \ref{itm:LinGauss} settings our \ac{lfsispf} accuracy surpasses that of the \ac{urpf} in all experiments (Figs. \ref{sfig:UR_avg} and \ref{sfig:UR_last}),  while operating with a similar number of \ac{fpm} and with a number of trainable parameters that is independent of trajectory length, {and without considering the measurements}.
On the more complex \ref{itm:NonLinGauss} (Figs. \ref{sfig:UR_avg_nonlin} and \ref{sfig:UR_last_nonlin}) and \ref{itm:UniformNoise}  (Figs. \ref{sfig:UR_avg_nongauss} and \ref{sfig:UR_last_nongauss}) cases, \ac{lf} outperforms the overall accuracy of the \ac{urpf} (Figs. \ref{sfig:UR_avg_nonlin} and \ref{sfig:UR_avg_nongauss}), but outperformed by it on the last time-step criterion (Figs. \ref{sfig:UR_last_nonlin} and \ref{sfig:UR_last_nongauss}). 
This can be attributed to the inferiority of the reference  \ac{pf} used in training, i.e.,  $SISPF\_Rs\_300$, as well as to the specialized training of the \ac{urpf}, that is trained separately for each specific trajectory and each specific task. 
This specialization may be the cause of the deteriorated accuracy of the \ac{urpf} on one task when trained for the other ($UrPF\_Rs\_\myVec{\xi}_2$ on the top sub-figures and $UrPF\_Rs\_\myVec{\xi}_1$ on the bottom sub-figures).

The \ac{lfurpf} experiments demonstrate the versatility of our algorithm, incorporating our \ac{lf} module to the \ac{dnn} augmented \ac{urpf} on selected time-steps. 
Even though the \ac{lf} module was trained to improve another (classical) \ac{pf}, its augmentation to the $UrPF\_\myVec{\xi}_2$ has improved it and the  $LF_{(-1)}-UrPF\_\myVec{\xi}_2$ accuracy on Figs. \ref{sfig:UR_avg_nonlin} and \ref{sfig:UR_avg_nongauss} surpasses those of $UrPF\_Rs\_\myVec{\xi}_1$ on the overall accuracy while maintaining accuracy on the last time-step on Figs. \ref{sfig:UR_last_nonlin} and \ref{sfig:UR_last_nongauss}.
Moreover, while the \ac{lf} module comes at the cost of some excessive complexity, as analyzed in Subsection~\ref{ssec:Discussion}, its gains in performance greatly outweigh its additional induced latency that is reported in  Table \ref{tab:stat_est_timings}. The runtime values, reported in Table \ref{tab:stat_est_timings}, together with Fig.~\ref{fig:LF_vs_Unrolling}, showcase the advantages of \ac{lf}, in allowing to notably enhance performance with limited particles without significantly increasing inference latency.

\begin{table}
\centering
    \fontsize{7.5pt}{10pt}\selectfont
    \begin{tabular}{|p{0.3cm}||p{1.1cm}|p{1.1cm}|p{1.1cm}|p{1.1cm}|p{1.1cm}|}
    \cline{1-6}
    & \ac{sispf} & \ac{sispf} & \ac{lfsispf} & \ac{urpf} & \ac{lfurpf} \\
    \hhline{|=|=|=|=|=|=|}
    $N$ & 300 & 25 & 25 & 25 & 25 \\
    \hline
    mS & 335.499 & 31.323 & 72.183 & 177.197 & 213.320 \\
    \hline
    \end{tabular}
    \caption{Average tracking latency in milliseconds over $\kappa=12$ time-steps trajectories with $N_{\rm th}=0$. }
    \label{tab:stat_est_timings}
\end{table}

\subsection{Radar Target Tracking}
\label{ssec:RadarTargetTracking} 
We proceed to evaluating \ac{lfpf} for  non-linear radar tracking, considering both a single and multiple targets. We use the non-linear settings outlined in~\cite[Sec. VI]{ubeda2017adaptive}, which allow us to compare the performance and latency of \acp{pf} to their \ac{lf} augmented versions in a scenario of practical importance.

\subsubsection{Experimental Setup}
In the considered scenarios, which are based on \cite[Sec. VI]{ubeda2017adaptive}, the state $\myVec{x}^k$ represents the positions and velocities of a fixed number of  targets $t$  in two-dimensional space over trajectories of length $\kappa=100$, with up to $t=10$ targets. Each target is independent of other targets and follows \ref{eqn:UrMotionModel} with $\phi(\myVec{x})=\myVec{x}$ and $A$ and $v^k$ taken from~\cite[Sec. VI.A]{ubeda2017adaptive}.
The measurements $\myVec{z}^k$ capture a {$13 \times 13$} sensor response solely to the targets' locations, following the sensor model of~\cite[Sec. VI.A]{ubeda2017adaptive}. We consider three different settings:
 \begin{enumerate}[label={Y\arabic*}]
    \item \label{itm:TargSingleCalib} Single target in {\em calibrated} settings, tracking with sensors located $10$ meters apart on $120\times120$ meters 2-dimensional plane, with different \acp{snr}.
    \item \label{itm:TargSingleMis}  Single target in {\em mismatched} settings, same assumed model where the actual sensors locations are perturbed with random Gaussian noise, with different \acp{snr}.
    \item \label{itm:TargMtt} \ac{mtt} of $1-10$ targets, a single \ac{snr}, {\em calibrated} settings.
\end{enumerate}

\subsubsection{Estimation Algorithms}
Similar to Algorithm~\ref{alg:LF_aug_PF},  we augment the \ac{apppf}, a version of the \ac{apf} proposed in \cite{ubeda2017adaptive}, prior to its integrated  Kalman filter that tracks the target velocity. Accordingly, we compare the following \acp{pf}:
\begin{itemize}
    \item \ac{apppf} from  \cite{ubeda2017adaptive} with single and multiple target support.
    \item \ac{lfapppf} for single target.  \ac{lf} module with $P=32$, $E=1$,  $B=1$; while $J,S=1$ (for  {\em calibrated} settings) and $J,S=2$ (for  {\em mismatched} settings).
    \item \ac{mtt} \ac{lfapppf}, with \ac{lf} module with $P=64$, $B=3$, $E=2$, $J=2$,  and $S=1$.
\end{itemize}
All \ac{lfapppf} benchmarks were trained according to Algorithm~\ref{alg:Train}, via \ref{itm:opt1} in the  {\em calibrated}  settings and via \ref{itm:opt2} in the  {\em mismatched} settings. 
All training was done on a single \ac{snr} using $N=100$ particles and $\tilde{N}=5000$ with batches and sub-batches sizes between $50-250$ and $5-25$.  Training, validation and testing performances are the average \ac{ospa}~\cite{schuhmacher2008consistent} metric (assigned per time-step), evaluated over all $\kappa=100$ time-steps, with cutoff of $10$ (and infinite cutoff on training) and order of $2$ (see ~\cite{schuhmacher2008consistent}  for details). 
The heatmap loss was computed using {\em staged meshgrid} with $L=5$ and $120\times120$ (on \ref{itm:TargSingleCalib},\ref{itm:TargSingleMis}) and $160\times160$ (on \ref{itm:TargMtt}) grid points. 
For all settings, we randomly combine a set of $n_t = 10^5$ single target sub-trajectories for training and two such sets of $n_t = 10^4$, for validation and testing. 



\subsubsection{Results} 
we divide our results into three parts with respect to  \ref{itm:TargSingleCalib}-\ref{itm:TargMtt} and compare the benchmarks for accuracy, {performance}, and latency, with accuracy results averaged over $10^4$ trajectories. We  visualize a single target trajectory tracking on \ref{itm:TargSingleMis} as well as an \ac{mtt} one in \ref{itm:TargMtt} settings with $t=10$ targets. As in Subsection~\ref{ssec:StateEstimation}, we also visualize the ability of  \ac{lf}  to enhance the particles and weights and their associated reconstructed \ac{pdf} on a single time-step on \ac{mtt}  with $t=3$. 

\begin{figure}
    \centering
        \includegraphics[width=\columnwidth]{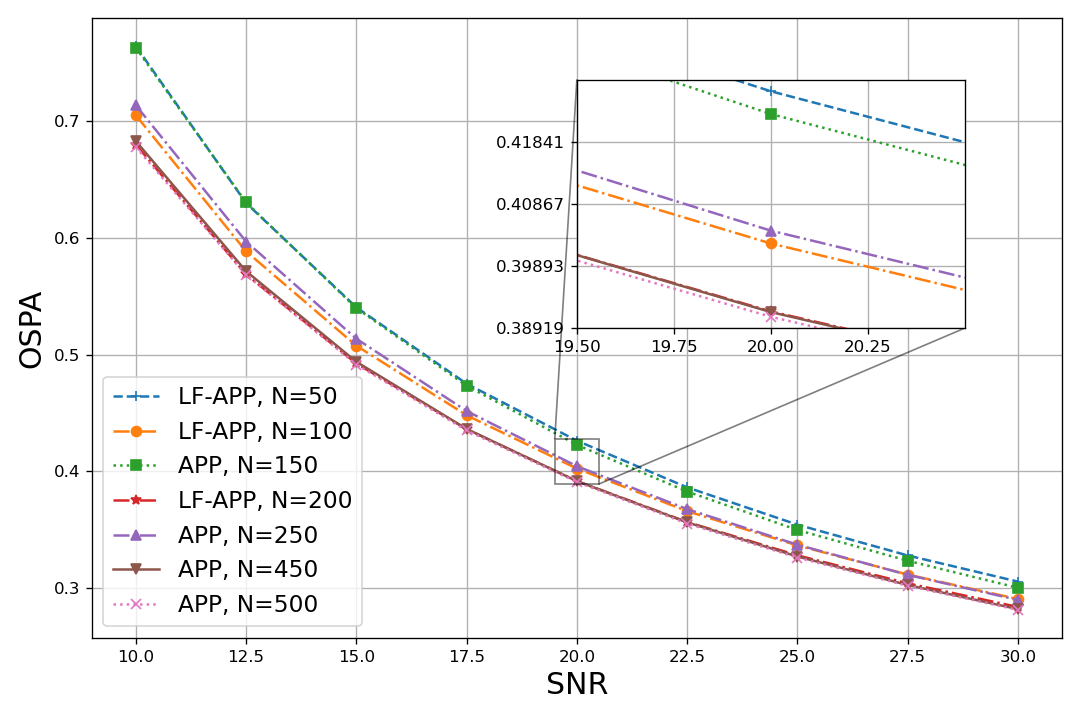}
    \caption{\ac{ospa} vs. \ac{snr} of \ac{apppf} and \ac{lfapppf} with different numbers of particles, (\ref{itm:TargSingleCalib}) experiment, {\em calibrated} settings.}
    \label{fig:accurate_ospa_vs_snr}
\end{figure}

\smallskip
{\bf Single target in {\em calibrated} settings (\ref{itm:TargSingleCalib}):}
We first compare the accuracy achieved by the \ac{apppf} to that of the \ac{lfapppf}, trained on ${\rm SNR}=20$. The resulting \ac{ospa} versus \ac{snr}, reported in Fig.~\ref{fig:accurate_ospa_vs_snr}, demonstrate that, although trained with $N=100$ particles and a single \ac{snr}, the correction term added by our \ac{lf} block allows \ac{apppf} to systematically achieve performance equivalent to $2-3$ times more particles on a wide rage of \acp{snr}. We also note that the improvement, induced by our Algorithm is more significant on lower \acp{snr}.

\begin{figure}
    \centering
        \includegraphics[width=\columnwidth]{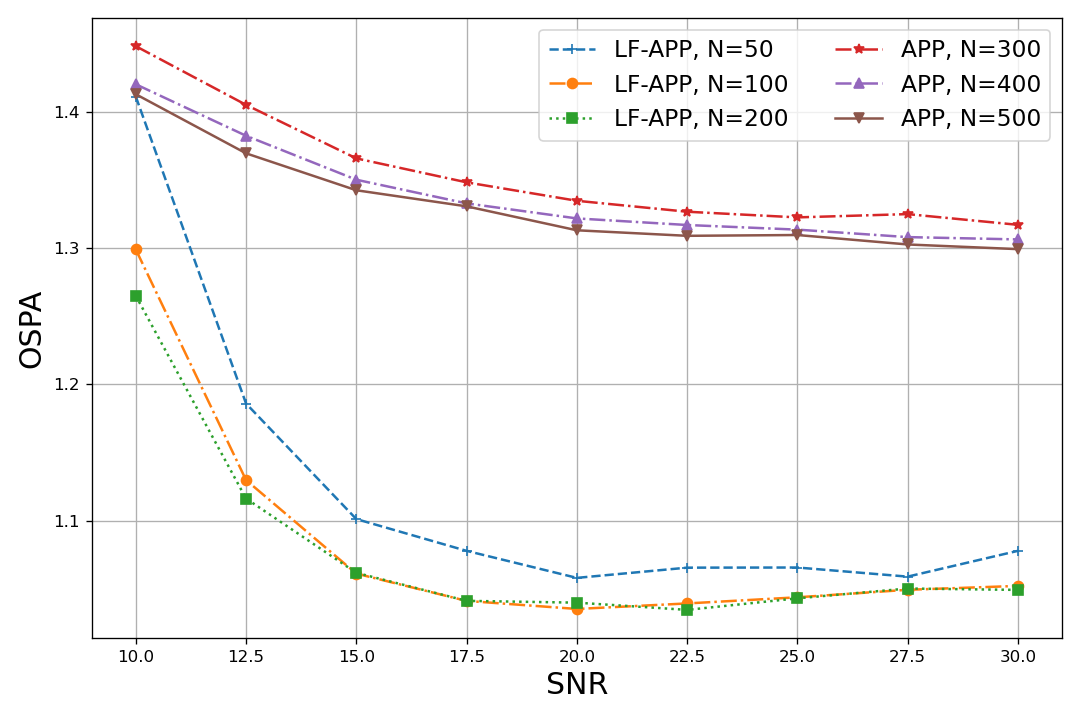}
    \caption{\ac{ospa} vs. \ac{snr} of \ac{apppf} and \ac{lfapppf} with different numbers of particles, (\ref{itm:TargSingleMis}) experiment, {\em mismatched} settings.}
    \label{fig:inaccurate_ospa_vs_snr}
\end{figure}

\begin{figure}
\vspace*{-4mm}
\centering
\begin{subfigure}[t]{.49\columnwidth}
    \centering
    \includegraphics[width=\textwidth]{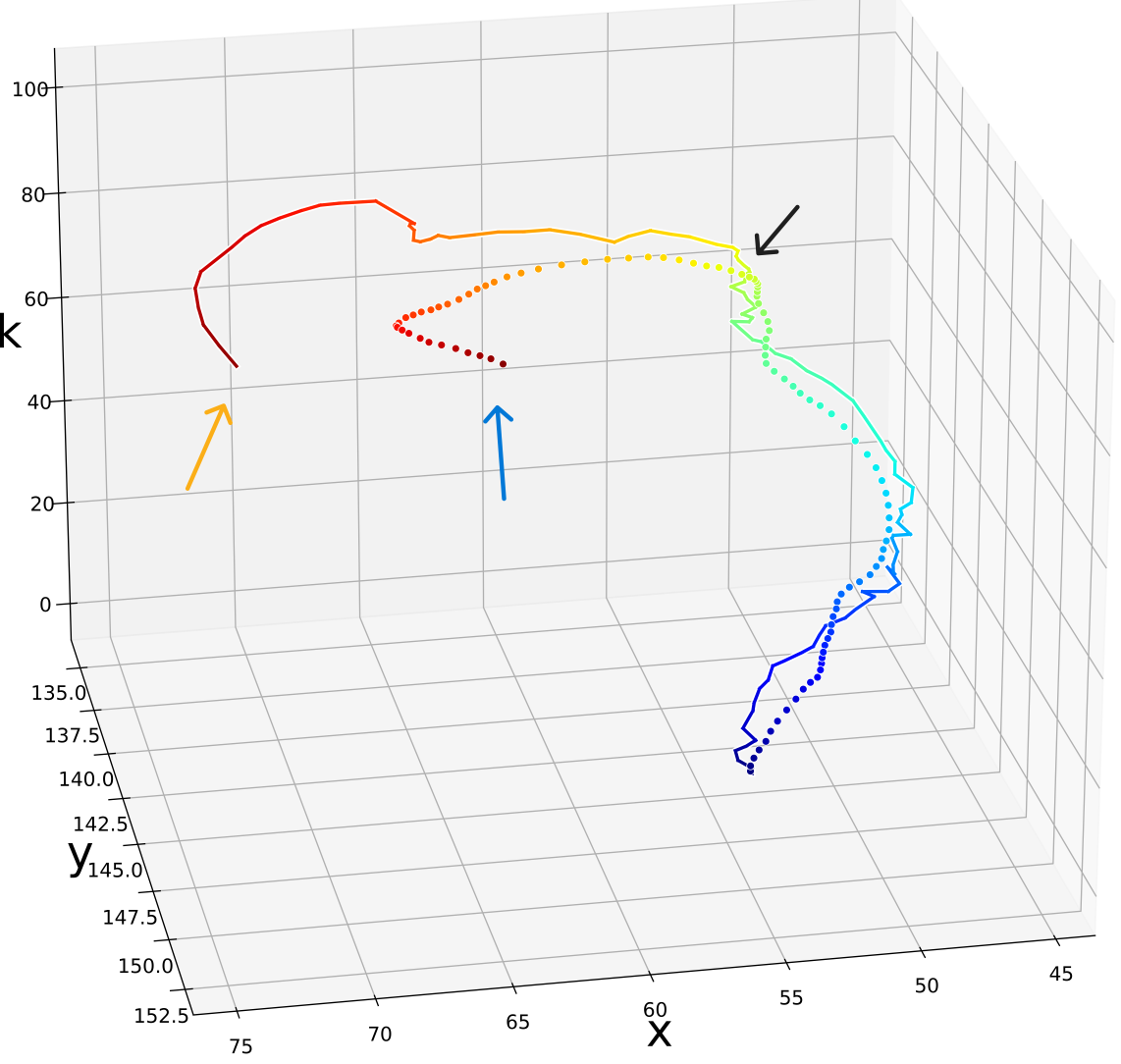}  
    \caption[]{\ac{apppf}, \ac{ospa}=$2.46$.}
    \label{sfig:tracking_traj_APP_single_targ_mismatched}
\end{subfigure}
\hfill
\begin{subfigure}[t]{.49\columnwidth}
    \centering
    \includegraphics[width=\textwidth]{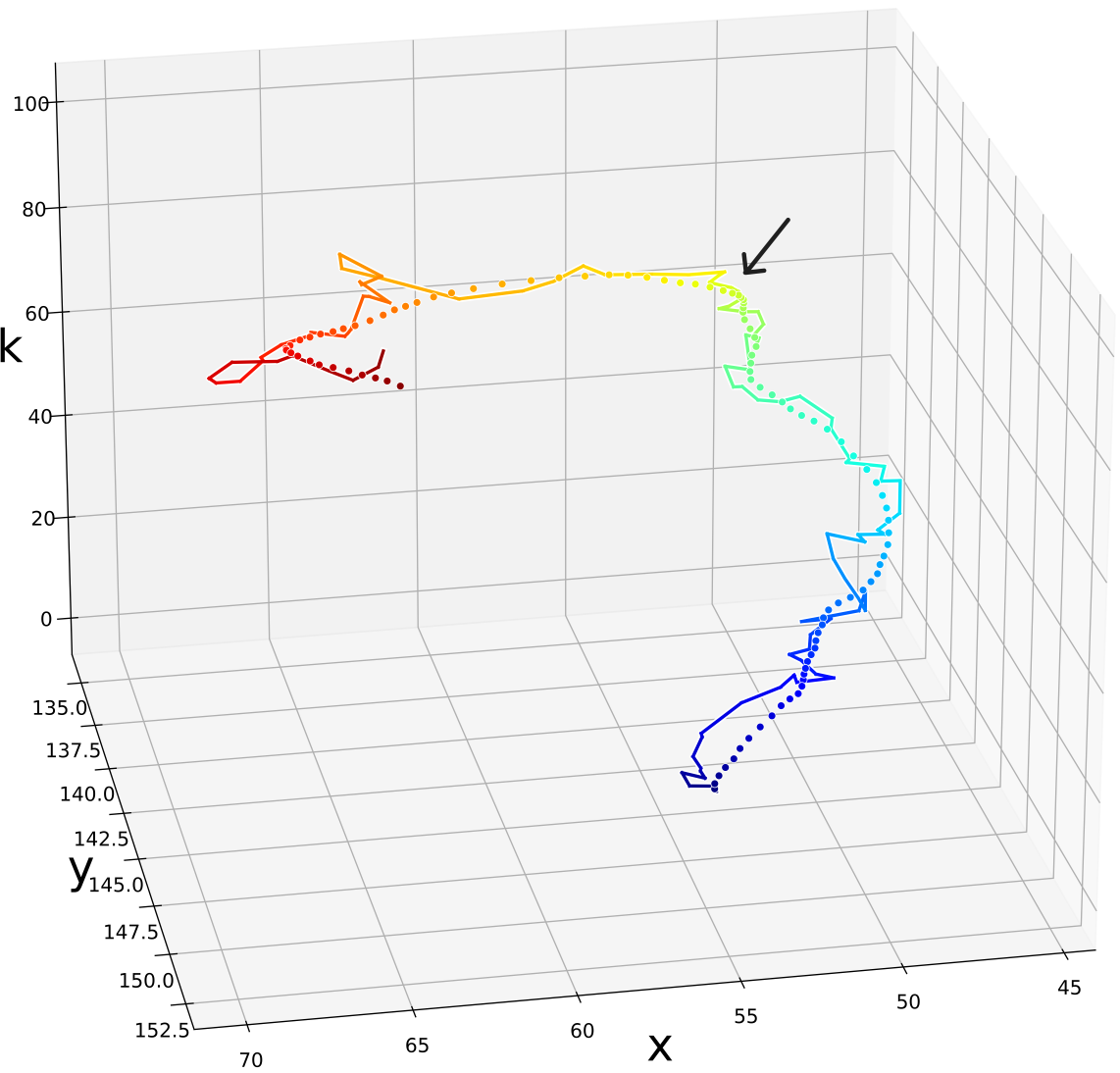}  
    \caption[]{\ac{lfapppf}, \ac{ospa}=$1.07$.}
        \vspace*{-2mm}
    \label{sfig:tracking_traj_LF_single_targ_mismatched}
\end{subfigure}
\vspace*{+2mm}
\caption{Radar single target simulation example with mismatched observation modelling on $\ac{snr}=10$ using the \ac{apppf} (\ref{sfig:tracking_traj_APP_single_targ_mismatched}) and the \ac{lfapppf} (\ref{sfig:tracking_traj_LF_single_targ_mismatched}). Tracking is done on $[x,y]$ plane over $\kappa=100$ time-steps shown as $k$ axis, using $N=100$ particles. True targets trajectories are shown as circles and reconstructed trajectories are shown as solid lines. Both trajectories colors represent the time-step, and arrows highlight tracking gaps. \ac{ospa} is computed with  cutoff=$\infty$.  
}
\vspace*{+3mm}

\label{fig:mismatched_single_target_tracking_illustration}
\end{figure}

\smallskip
{\bf Single target in {\em mismatched} settings (\ref{itm:TargSingleMis}):}
For the mismatched case, we employ \ac{lfapppf} with $J,S=2$, trained on ${\rm SNR}=10$. 
The results, reported in Fig.~\ref{fig:inaccurate_ospa_vs_snr} show that \ac{lfapppf}  notably improves performance with similar consistent improvement with increased $N$, but also yields an error floor for the model-based \ac{apppf}. To show that this gain is directly translated into improved tracking, we illustrate in Fig.~\ref{fig:mismatched_single_target_tracking_illustration} the tracking of a single   sub-state trajectory, where the \ac{lf} augmentation is  shown to notably improve tracking accuracy.

We proceed by evaluating the excessive latency of \ac{lfapppf}. We report in Table~\ref{tab:dnn1_time2} the latency of both \ac{apppf} and \ac{lfapppf} used for both experiments in milliseconds. The timing results in Table~\ref{tab:dnn1_time2} indicate that while  latency grows dramatically with $N$, the excessive latency induced by incorporating our \ac{lf} block is minor. Combining this with  Figs.~\ref{fig:accurate_ospa_vs_snr} and \ref{fig:inaccurate_ospa_vs_snr} showcases the ability of \ac{lfapppf} in allowing \acp{pf} (such as the \ac{apppf}) to  meet \ref{itm:dist}-\ref{itm:latency} on a wide range of single target  scenarios.

\begin{table}
\centering
    \fontsize{7.5pt}{10pt}\selectfont
    \begin{tabular}{|p{0.6cm}||p{1.8cm}|p{1.8cm}|p{1.8cm}||}
    \cline{1-4}
    $N$ & \ac{apppf}& \ac{lfapppf}, $(J,S)=(1,1)$ & \ac{lfapppf}, $(J,S)=(2,2)$\\
    \hline  
    50  & 8.200 & 11.548 & 14.039 \\
    100 & 13.993 & 19.400 & 20.828 \\
    200 & 34.147 & 36.177 & 38.082 \\
    300 & 50.221 & 53.046 & 55.666 \\
    400 & 66.195 & 69.797 & 73.393 \\
    500 & 81.986 & 86.749 & 91.245 \\
    \hline              
    \end{tabular} 
    \caption{Average tracking latency per time-step in milliseconds for the single target benchmarks of  Subsection~\ref{ssec:RadarTargetTracking}.  Tested on $\kappa=100$ time instances long trajectories.}
    \label{tab:dnn1_time2}
\end{table}

\smallskip
{\bf \ac{mtt} in {\em calibrated} settings (\ref{itm:TargMtt}):}
We proceed to compare the \ac{apppf} and the \ac{lfapppf} for their accuracy, performance  and latency for a known and fixed number of targets $t$, taking values between $1$ and $10$ and with different numbers of particles $N$. Training was done by jointly learning~\cite{raviv2023adaptive} on data corresponding to $1$, $3$, $5$  (with equal probability) and $8$ (10\% of trajectories)  targets tracking scenarios with ${\rm SNR}=20$, and with validation on $t=4$ targets tracking. 
Since the basic form of our \ac{lf} tested here has a per time-step operation, and in order to isolate its evaluation, we address training and testing in the same manner, and the mapping between estimations and true targets is done separately on each time-step according to the best \ac{ospa}, effectively disregarding target swaps between time-steps. 

\begin{figure}
\vspace*{-4mm}
\centering
\begin{subfigure}[t]{.32\columnwidth}
    \centering
    \includegraphics[width=\textwidth]{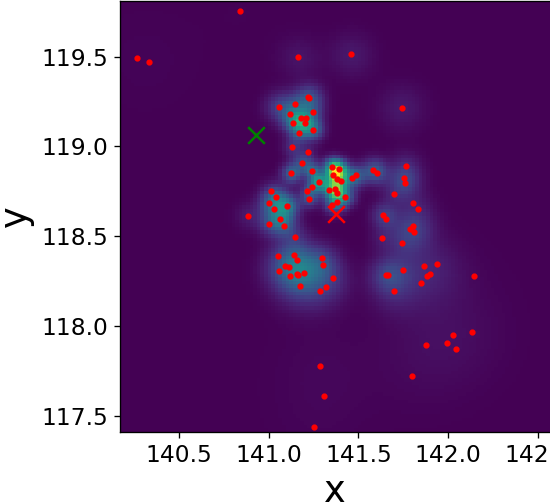}  
    \caption[]{\ac{lf} input $j=1$}
    \label{sfig:heatmap_staged_grid_before_nn3_1}
\end{subfigure}
\hfill
\begin{subfigure}[t]{.32\columnwidth}
    \centering
    \includegraphics[width=\textwidth]{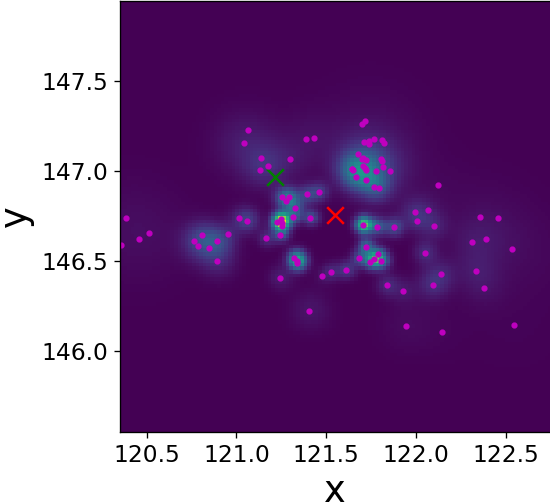}  
    \caption[]{\ac{lf} input $j=2$}
        \vspace*{-2mm}
    \label{sfig:heatmap_staged_grid_after_nn3_2}
\end{subfigure}
\begin{subfigure}[t]{.32\columnwidth}
    \centering
    \includegraphics[width=\textwidth]{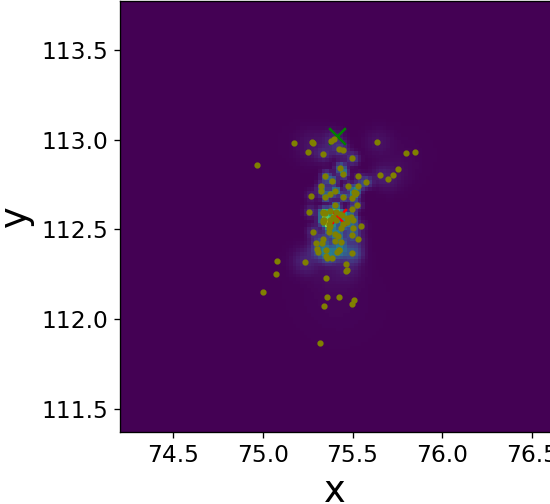}  
    \caption[]{\ac{lf} input $j=3$}
        \vspace*{-2mm}
    \label{sfig:heatmap_staged_grid_before_nn3_3}
\end{subfigure}
\vspace*{+5mm}

\begin{subfigure}[t]{.32\columnwidth}
    \centering
    \includegraphics[width=\textwidth]{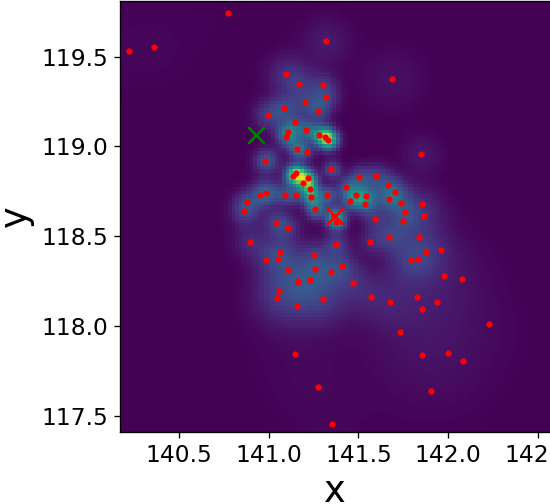}  
    \caption[]{\ac{lf} output $j=1$}
    \label{sfig:heatmap_staged_grid_after_nn3_1}
\end{subfigure}
\begin{subfigure}[t]{.32\columnwidth}
    \centering
    \includegraphics[width=\textwidth]{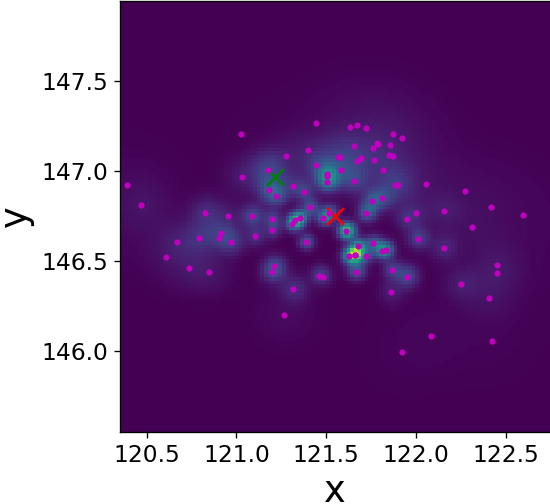}  
    \caption[]{\ac{lf} output $j=2$}
    \label{sfig:sorted_wts_staged_grid_after_2}
\end{subfigure}
\begin{subfigure}[t]{.32\columnwidth}
    \centering
    \includegraphics[width=\textwidth]{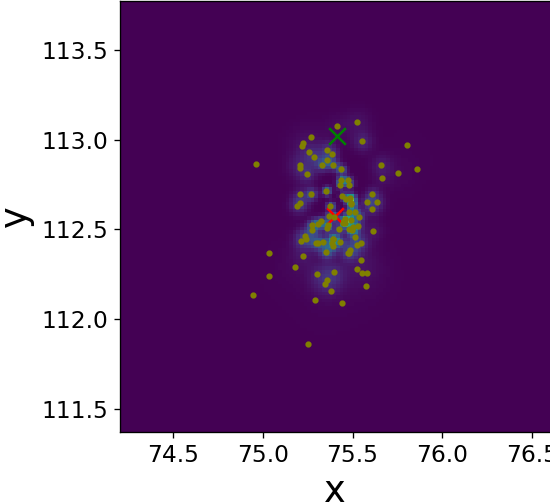}  
    \caption[]{\ac{lf} output $j=3$}
    \label{sfig:heatmap_staged_grid_after_nn3_3}
\end{subfigure}
\vspace*{+2mm}
\vskip\baselineskip

\begin{subfigure}[t]{.49\columnwidth}
    \centering
    \includegraphics[width=\textwidth]{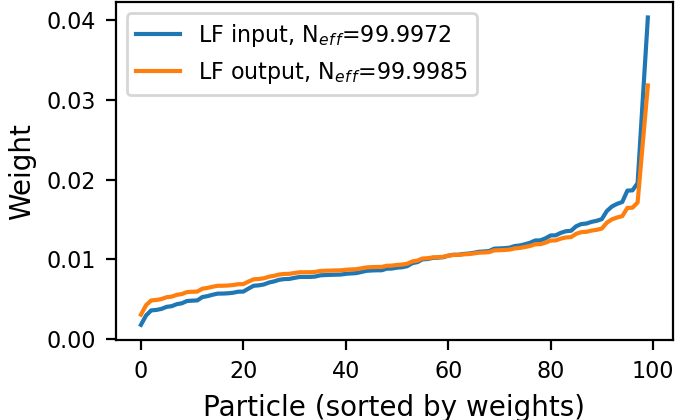}  
    \caption[]{Sorted weights.}
    \label{sfig:staged_grid_sorted_wts}
\end{subfigure}
\begin{subfigure}[t]{.49\columnwidth}
    \centering
    \includegraphics[width=\textwidth]{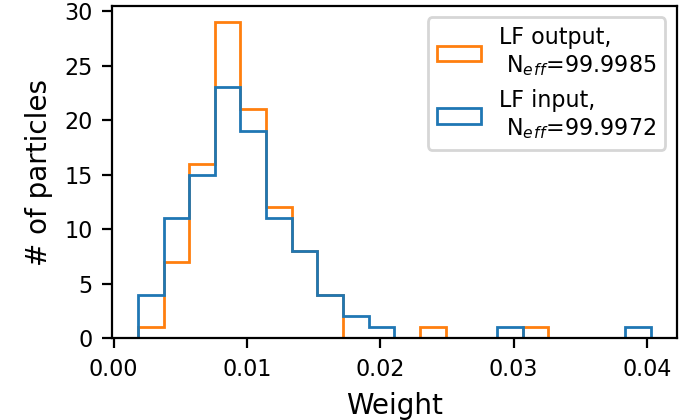}  
    \caption[]{Weights histograms.}
    \label{sfig:staged_grid_wts_hist}
\end{subfigure}
\vspace*{+3mm}
\caption{\ac{lf} \ac{mtt} sub-particles and weights adjustment example with three targets on the \ac{mtt} \ac{lfapppf} experiment.  
The \ac{lf} module input (Figs.~\ref{sfig:heatmap_staged_grid_before_nn3_1}-\ref{sfig:heatmap_staged_grid_before_nn3_3}) and output (Figs.~\ref{sfig:heatmap_staged_grid_after_nn3_1}-\ref{sfig:heatmap_staged_grid_after_nn3_3}) sub-particles and their respective induced \ac{pdf} as part of the  heatmap loss. 
The particles are marked with dots; the desired state and its estimate are marked with green and red crosses, respectively.
Fig.~\ref{sfig:staged_grid_sorted_wts} shows the sorted weights (common to all three sub-particles) and Fig.~\ref{sfig:staged_grid_sorted_wts} shows the histograms of the weights at the input and output of the \ac{lf} module. }
\label{fig:staged_grid_mtt_particles_spread}
\end{figure}

Fig.~\ref{fig:staged_grid_mtt_particles_spread} illustrates the operation of our \ac{lf} block on a single frame three target tracking on time-step $k=19$. Figs.~\ref{sfig:heatmap_staged_grid_before_nn3_1}-\ref{sfig:heatmap_staged_grid_after_nn3_3} visualize its improving of the spread of the three sets of sub-particles by dispersing clusters and reducing outliers.  Figs.~\ref{sfig:staged_grid_sorted_wts}-\ref{sfig:staged_grid_wts_hist} show that  our trained \ac{lf} module  encourages low variance of the weights without explicitly doing so in the loss, implying a better utilization of the particles. We again observe that $N_{\rm eff}$ is close to $N=100$ already at the  input of our \ac{lf} module, hinting the effectiveness of our \ac{lf} algorithm in describing the \ac{pdf} along the whole trajectory, as well as strengthens  its effectiveness ~\cite{sarkka2023bayesian}, as it is useful also for enhancing in settings where $N_{\rm eff}\approx N$.   

\begin{figure*}
\vspace*{-7mm}
\centering
\begin{subfigure}[t]{.99\columnwidth}
    \centering
    \includegraphics[width=\textwidth]{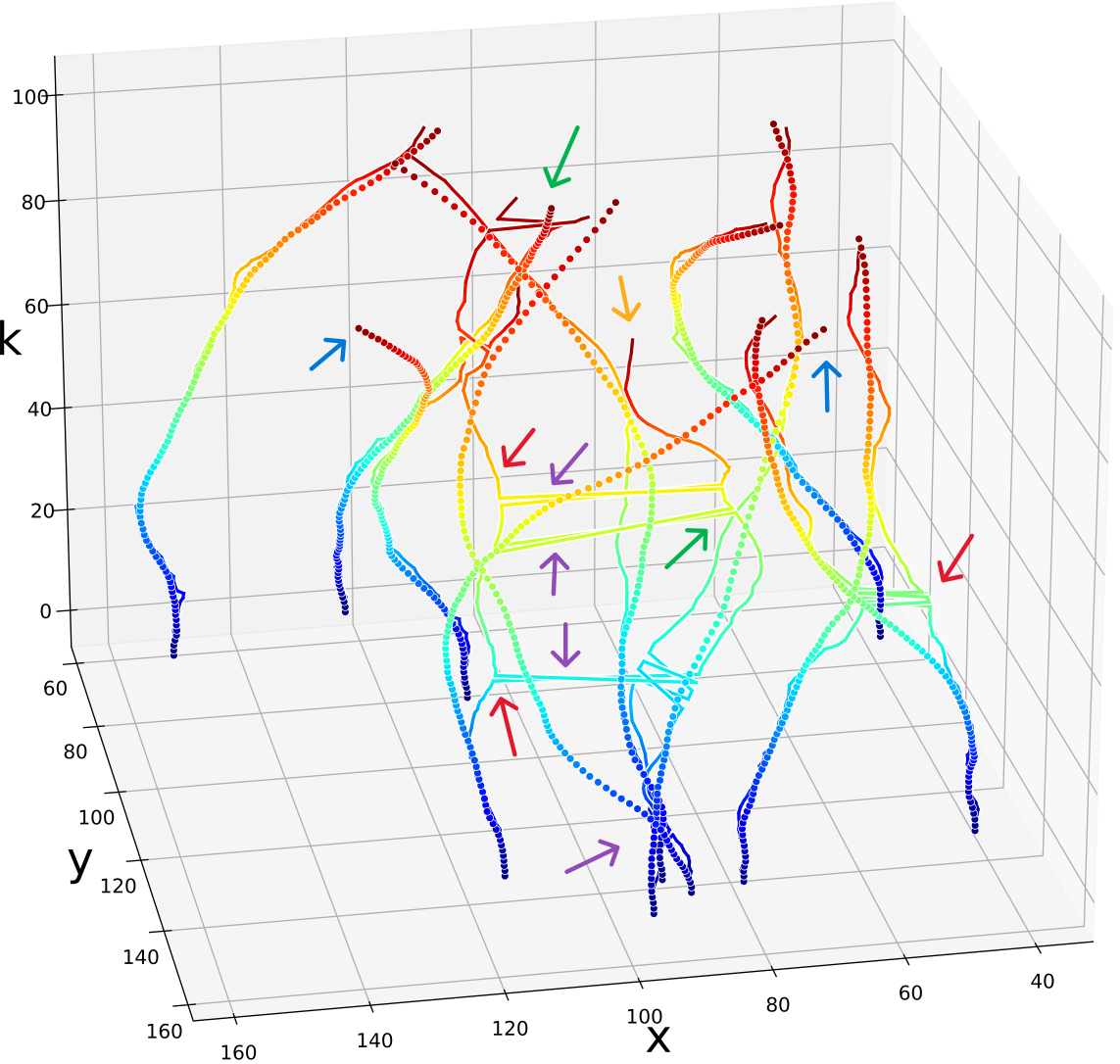}  
    \caption[]{\ac{apppf}, \ac{ospa}=$3.63$, and $7.36$ with cutoffs $10$ and $\infty$.}
    \label{sfig:traking_traj_APP_mtt}
\end{subfigure}
\begin{subfigure}[t]{.99\columnwidth}
    \centering
    \includegraphics[width=\textwidth]{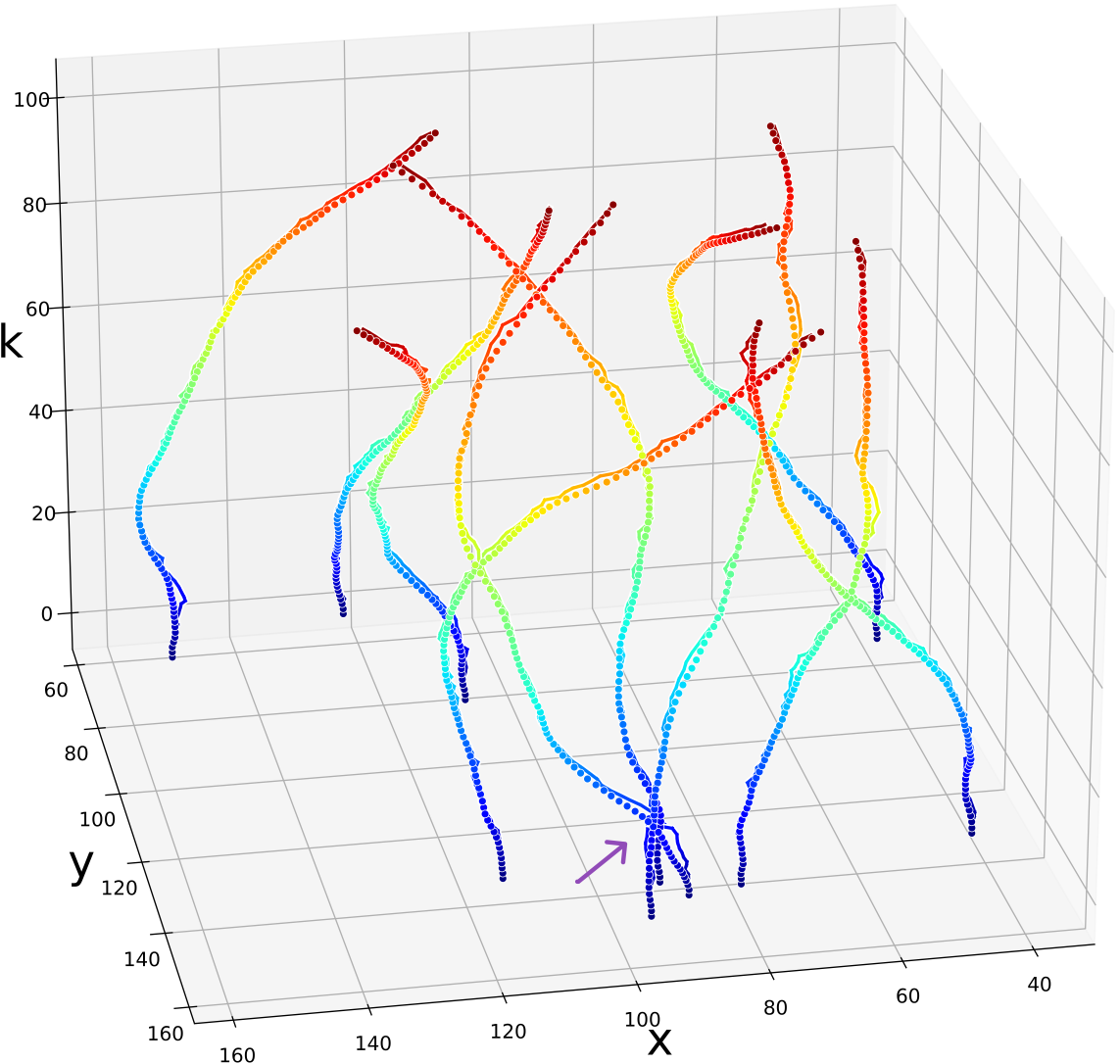}  
    \caption[]{\ac{lfapppf}, \ac{ospa}=$0.55$, with both cutoffs.}
        \vspace*{-2mm}
    \label{sfig:traking_traj_LF_mtt}
\end{subfigure}
\vspace*{+5mm}
\caption{Radar \ac{mtt} simulation example with $t=10$ targets using the \ac{apppf} (\ref{sfig:traking_traj_APP_mtt}) and the \ac{lfapppf} (\ref{sfig:traking_traj_LF_mtt}). Tracking is done on $[x,y]$ plane over $\kappa=100$ time-steps shown as $k$ axis, using $N=100$ particles. True targets trajectories and reconstructed ones are shown as circles and solid lines, respectively. Colors represent the time-step, and arrows highlight tracking gaps.}
\label{fig:mtt_tracking_illustration}
\end{figure*}

Single time-step improvements on the spread of the sub-particles as presented on Fig.~\ref{fig:staged_grid_mtt_particles_spread}, applied to each time-step, have a major effect on the overall tracking of the targets along the {\em full} state trajectory. 
Fig.~\ref{fig:mtt_tracking_illustration} illustrates this benefit on a $10$ target \ac{mtt} simulation. 
There, we can see the improvement in terms of a diminished number of lost targets, better targets-estimations mapping injectiveness and better accuracy. 
Further, even though we did not explicitly encourage it in our loss function, we also witness less targets-estimations mapping swaps. 
Examples of the aforementioned improvements are highlighted on Fig.~\ref{fig:mtt_tracking_illustration} with arrows of respective colors.

\begin{figure}
\vspace*{-2mm}
    \centering
        \includegraphics[width=\columnwidth]{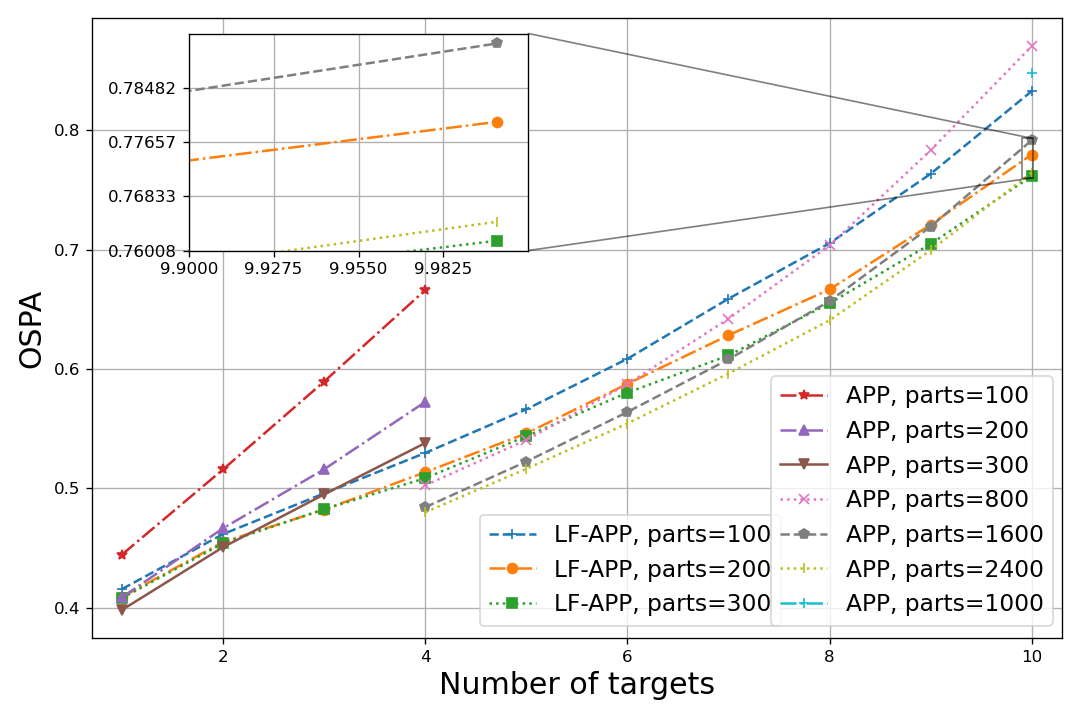}
    \caption{\ac{ospa} vs. number of targets $t$, with different numbers of particles, \ac{apppf} and \ac{lfapppf}, on the \ac{mtt} (\ref{itm:TargMtt}) experiment.}
    \label{fig:mtt_ospa_vs_snr}
\end{figure}

The translation of these improvements into tracking accuracy is showcased on  Fig.~\ref{fig:mtt_ospa_vs_snr}.  
rained with $100$ particles and applied with varying number of particles, Fig.~\ref{fig:mtt_ospa_vs_snr} illustrates that \ac{lf} provides greater improvement in performance as the number of targets increases. While trained with up to $8$ targets on only $10\%$ of the trajectories,  \ac{lf} achieves a significant reduction in the required number of particles when handling $t=10$ targets, consistently reducing it by a notable factor of $8\times$, and up to $10\times$ with a $100$ particles.

Complementary to Fig.~\ref{fig:mtt_ospa_vs_snr},  Table \ref{tab:mttdnn1_time2} compares the tracking  latency of the \ac{apppf} and \ac{lfapppf} with different numbers of particles and targets. Combining the results in Fig.~\ref{fig:mtt_ospa_vs_snr} and Table \ref{tab:mttdnn1_time2}, and comparing them to the \ref{itm:TargSingleCalib}   results in Fig.~\ref{fig:accurate_ospa_vs_snr} and Table \ref{tab:dnn1_time2}, consistently demonstrate that the benefit of our \ac{lf} algorithm in coping with the challenging task of tracking multiple targets with limited latency is  more prominent and potentially more beneficial than in single target settings.

\newcolumntype{g}{>{\columncolor{Gray}}c}

\begin{table}
\centering
    \fontsize{7.5pt}{10pt}\selectfont
    \begin{tabular}{|p{0.4cm}|p{0.8cm}||p{0.68cm}|p{0.68cm}|p{0.68cm}|p{0.68cm}|p{0.68cm}|p{0.68cm}|p{0.68cm}||}
    \cline{1-8}
    \multicolumn{2}{|c|}{\backslashbox[20mm]{$N$}{$t$}} & 1 & 2 & 4 & 6 & 8 & 10 \\
    \hhline{|=|=|=|=|=|=|=|=|}
    \multirow{ 2}{*}{100} & \ac{apppf} & 14.62 & 28.24 & 49.34 & 73.32 & 101.393 & 135.10 \\
    \hhline{~|-|-|-|-|-|-|-|}
    & \cellcolor{lightgray} \ac{lfapppf}  & \cellcolor{lightgray} 24.08 & \cellcolor{lightgray} 41.33  & \cellcolor{lightgray} 66.24 & \cellcolor{lightgray} 98.32 & \cellcolor{lightgray} 135.52 & \cellcolor{lightgray} 172.35 \\
    \hline
    \multirow{ 2}{*}{200}  & \ac{apppf} & 35.52 & 53.89 & 95.29 & 142.35 & 198.011 & 260.86 \\
    \hhline{~|-|-|-|-|-|-|-|}
    & \cellcolor{lightgray} \ac{lfapppf}  & \cellcolor{lightgray} 44.66 & \cellcolor{lightgray} 79.95  & \cellcolor{lightgray} 127.64 & \cellcolor{lightgray} 189.11 & \cellcolor{lightgray} 263.19  & \cellcolor{lightgray} 337.53 \\
    \hline
    \multirow{ 2}{*}{300}  & \ac{apppf} & 52.12 & 79.51 & 141.14 & 211.19 & 295.751 & 386.41 \\
    \hhline{~|-|-|-|-|-|-|-|}
    & \cellcolor{lightgray}  \ac{lfapppf} & \cellcolor{lightgray} 64.94   & \cellcolor{lightgray} 116.26 & \cellcolor{lightgray} 189.63 & \cellcolor{lightgray} 288.43 & \cellcolor{lightgray} 391.55 & \cellcolor{lightgray} 505.67 \\
    \hline
    \multirow{ 1}{*}{800}  & \ac{apppf} & 129.93 & 203.29 & 362.40 & 542.73 & 803.78 & 999.30  \\
    \hline
    \multirow{ 1}{*}{1600}  & \ac{apppf} & 257.77 & 404.08 & 720.87 & 1087.01 & 1624.70 & 1996.31 \\
    \hline
    \multirow{ 1}{*}{2400}    & \ac{apppf} & 385.63 & 595.86 & 1046.25 & 1569.93 & 2184.01 & 2824.22 \\
    \hline
    \end{tabular}
    \caption{Average tracking latency per time-step in milliseconds for the study of Fig.~\ref{fig:mtt_ospa_vs_snr}. Tested on $\kappa=100$  long trajectories.}
    \label{tab:mttdnn1_time2}
\end{table}

\section{Conclusions}
\label{sec:conclusions}

We introduced a neural augmentation of \acp{pf} using a novel concept of \ac{lf}, that supports {\em changing} numbers of  sub-states  and  particles. 
\ac{lf} is designed to enhance state estimation by adjusting a \ac{pf}'s particles at any point in its flow, {\em solely} based on the interplay between the particles, resulting in an improved output.  
We realize \ac{lf} using a dedicated \ac{dnn} architecture that accounts for the permutation equivariance property of the particle-weight pairs in \acp{pf}, and introduce a dedicated training  framework that supports both supervised and {\em unsupervised} learning, while accommodating  high-dimensional  sub-state tracking and with {\em indifference} to the particles acquirement procedure. 
Once trained, the \ac{lf} module is  easily {\em transferable} to most \acp{pf}, and can even be combined with alternative \ac{dnn} \ac{pf} augmentations. 
We experimentally exemplified the gains of \ac{lf} in terms of performance, latency, and robustness and even in overcoming mismatched modelling, demonstrating its potential for enhancing variety of \acp{pf} and as a foundation for enhancement of more intricate \ac{pf} filtering tasks.

\bibliographystyle{IEEEtran}
\bibliography{IEEEabrv,refs}

\end{document}